\documentclass[preprint,showpacs,preprintnumbers,amsmath,amssymb]{revtex4}

\usepackage{graphicx}
\usepackage{dcolumn}
\usepackage{bm}

\newcommand{\comment}[1]{}
\newcommand{\bea}{\begin{eqnarray}}
\newcommand{\eea}{\end{eqnarray}}

\begin{document}

\title[Classification of equation]
{On the complete Lie point symmetries classification of the mixed quadratic-linear Li$\acute{\textbf{e}}$nard type equation $\ddot{x}+f(x)\dot{x}^2+g(x)\dot{x}+h(x)=0$}

\author{Ajey K. Tiwari$^1$}%
\author{S. N. Pandey$^1$}%
\email{snp@mnnit.ac.in (S. N. Pandey)}
\author{M. Senthilvelan$^2$}%
\email{ velan@cnld.bdu.ac.in (M. Senthilvelan)}
\author{M. Lakshmanan$^2$}%
\email{ lakshman@cnld.bdu.ac.in (M. Lakshmanan)}
\affiliation{%
$^1$ Department of Physics, Motilal Nehru National Institute of
Technology, Allahabad - 211 004, India\\$^2$ Centre for Nonlinear Dynamics, School of Physics, Bharathidasan University,
 Tiruchirapalli - 620 024, India}%
\date{\today} 

\begin{abstract}
In this paper we develop a systematic and self consistent procedure based on a set of compatibility conditions for identifying all maximal (eight parameter) and non-maximal (one and two parameter) symmetry groups associated with the mixed quadratic-linear Li$\acute{e}$nard type equation, $\ddot {x} + f(x){\dot {x}}^{2} + g(x)\dot{x}+h(x)= 0$, where $f(x),\,g(x)$ and $h(x)$ are arbitrary functions of $x$. With the help of this procedure we show that a symmetry function $b(t)$ is zero for non-maximal cases whereas it is not so for the maximal case. On the basis of this result the symmetry analysis gets divided into two cases, $(i)$ the maximal symmetry group $(b\neq0)$ and $(ii)$ non-maximal symmetry groups $(b=0)$. We then identify the most general form of the mixed-quadratic linear Li$\acute{e}$nard type equation in each of these cases. In the case of eight parameter symmetry group, the identified general equation becomes linearizable. We present a  specific example of physical interest. In the case of non-maximal symmetry groups the identified equations are all integrable. The integrability of all the equations is proved either by providing the general solution or by constructing time independent Hamiltonians. We also analyse the underlying equivalence transformations.
\end{abstract}

\maketitle
\section{Introduction}
\label{sec1}
\subsection{Motivation}
Ordinary differential equations (ODEs) play a crucial role in all areas of science and technology because they help in understanding diverse physical 
phenomena. Hence different aspects of ODEs have been studied over the past three centuries or so \cite{baum,ibra,blum,olver,step,euler,hill,hyd,cant,lak}. 
Inspired by the work of Abel and Galois, Sophus Lie \cite{baum,ibra,blum} was interested in developing a general theory for the integration of ODEs. 
He started a new area which is now called as the symmetry analysis of differential equations in which he investigated the continuous groups of 
transformations that leave the differential equations invariant and showed that all the integration methods for ODEs can be obtained from his theory. 
Lie gave a classification of ODEs in terms of their symmetry groups, thereby identified the full set of equations which can be integrated or reduced to 
lower order equations by his method. Since then several contributions have been made on symmetry group classification of ODEs 
\cite{olver,step,euler,baum,ibra,hill,hyd,cant,blum,lak,maho,maho1,leach,leach2,govinder,gat}. Over the years several generalization of classical Lie 
method have also been proposed in order to explore more generalized symmetries to establish the integrability of certain nonlinear dynamical systems. 
Notable extensions/algorithms constructed in this direction are $(i)$ contact symmetries, $(ii)$ potential symmetries, $(iii)$ $\lambda-$symmetries and 
$(iv)$ telescopic symmetries and so on \cite{blum,olver,step,euler,hill,hyd,muri1,pucci}. In this paper we consider only Lie point symmetries associated with the given nonlinear ODEs.   

Recently, three of the present authors and Bindu \cite{snp1} have studied the Lie point symmetries of a nonlinear Li$\acute{e}$nard type equation with a linear velocity $(\dot{x})$ term,
\begin{equation}
\ddot {x} + f(x){\dot {x}} + g(x)= 0,\,\,\,(\cdot=\frac{d}{dt}),\label{intr1}
\end{equation}
where $f(x)$ and $g(x)$ are arbitrary smooth functions of $x$ and over dot denotes differentiation with respect to $t$, and identified several  interesting integrable and linearizable equations. In their two part analysis, the authors have isolated equations that admit one and two parameter Lie point symmetries, in the first part, and identified equations that admit maximal (eight parameter) Lie point symmetries symmetries in the second part. They have also proved the integrability of all the equations obtained in their analysis either by providing the general solution or by constructing a time independent Hamiltonian (Liouville integrable). The linearizing transformations and the general solutions for all the identified linearizable nonlinear ODEs in this class have also been given. 

Very recently, the present authors \cite{akt} have carried out Lie symmetry analysis of a quadratic (in $\dot{x}$) Li$\acute{\textbf{e}}$nard type equation 
\begin{equation}
\ddot {x} + f(x){\dot {x}}^2 + g(x)= 0,\label{intr2}
\end{equation}
where $f(x)$ and $g(x)$ are arbitrary smooth functions of $x$. We have identified and classified all the specific equations belonging to the class (\ref{intr2}) which admit one, two, three and eight parameter symmetry groups and explored the dynamics associated with them. We mention here that certain equations which we have identified as linearizable/integrable through Lie symmetry analysis have also been investigated in other perspectives. For example, the modified Emden equation (MEE) with linear external forcing, $\ddot{x}+3x\dot{x}+x^3+\lambda_1x=0$, has been studied in detail and shown to admit certain remarkable properties, including $(i)$ amplitude independent frequency of oscillations and $(ii)$ non-standard Lagrangian/Hamiltonian description. Besides the above, Lagrangian multipliers, integrating factors, $\lambda-$symmetries, exponential nonlocal symmetries and Darboux polynomials have also been explored for this nonlinear ODE \cite{vkc11,vkc1,vkc12,bru}. The isochronous systems that admit Lie point symmetries in the class (\ref{intr2}) have also been considered earlier in the literature \cite{chou,saba}. Some notable equation that belongs to class (\ref{intr2}) are $(i)$ Mathews-Lakshmanan (ML) oscillator, $\ddot{x}-\frac{\lambda \,x}{1+\lambda \,x^2}\dot{x}^2+\frac{\omega_0^2\,x}{1+\lambda \,x^2}=0$ and $(ii)$ an isochronous oscillator equation, $\ddot{x}+\frac{3ax}{1-ax^2}\dot{x}^2+x-ax^3=0$ and so on. Interestingly the ML oscillator, eventhough it admits only one Lie point symmetry it has been proved to be linearizable through nonlocal transformations. Due to its unusual property the underlying equation has been studied by many authors at the classical as well as quantum levels.

A wide range of investigations carried on certain ODEs, such as the generalized MEE equation and Mathews-Lakshmanan (ML) oscillator equation which are captured by group invariance properties motivates us to classify/identify linearizable and integrable nonlinear ODEs of a general mixed quadratic-linear (in $\dot{x}$) Li$\acute{e}$nard type equation \cite{chou,saba}
\begin{eqnarray}
A(\ddot{x},\dot{x},x)\equiv \ddot{x}+f(x)\dot{x}^2+g(x)\dot{x}+h(x)=0,\label{z1}
\end{eqnarray}
where $f(x),\,g(x)$ and $h(x)$ are arbitrary functions of $x$ and overdot denotes differentiation with respect to $t$, which is much more challenging than the study of (\ref{intr1}) and (\ref{intr2}). One can observe that (\ref{intr1}) and (\ref{intr2}) are subcases of (\ref{z1}). Infact, Eq. (\ref{z1}) distinguishes itself by admitting several physically important systems. Lotka-Voltera equation (written as a second order ODE), second order Gambier equation, when the coefficients are assumed to be constant parameters, and second order Ricatti equation are notable examples of this class of equation. In this paper, by developing a rather general approach, we systematically identify and classify all the equations, belonging to class (\ref{z1}), which admit one, two and eight parameter Lie point symmetry groups, from a group theoretic point of view and explore the dynamics associated with them. We also show that there exists no ODE which admits three parameter Lie point symmetry group, when both $f(x)$ and $g(x)$ are nonzero in (\ref{z1}).

\subsection{Outcome}
After performing an in-depth analysis we conclude that the general form of linearizable equation belonging to the class (\ref{z1}) should 
be of the form
\begin{eqnarray}\label{intr3}
\ddot{x}+f(x){\dot{x}}^2+\left(k_1\int{e^{\int{f(x)dx}}dx} +k_2\right)\dot{x}+e^{-\int{f(x)dx}}\left[\frac{k_1^2}{9}\left(\int{e^{\int{f(x)dx}}dx}\right)^3\right.\nonumber\\
\left.+\frac{k_1k_2}{3}\left(\int{e^{\int{f(x)dx}}dx}\right)^2+k_3\,\left(\int{e^{\int{f(x)dx}}dx}\right)+k_4\right]=0,\label{z2.1}
\end{eqnarray}
where $k_1,k_2,k_3$ and $k_4$ are constant parameters. We not only present the eight Lie point symmetries of this equation but also report the associated two time dependent integrals of motion and the invertible linearizing point transformation for (\ref{z2.1}) which takes the later into free particle equation, from which the general solution is also obtained. The linearizing transformation which we report here is new to the literature as far as we are aware of.  We note here that the Lie point symmetries of Eq. (\ref{z2.1}) are in a complicated form (vide (\ref{sym049})) below. Since it is very difficult to construct the linearizing transformations from the Lie point symmetries itself we employ another method which was advocated recently by Chandrasekar \textit{et al} which in turn provides the linearizing transformation in a very straightforward manner \cite{vkc1}.   

We then focus our attention to lesser parameter symmetry groups, that is, one and two parameter Lie point symmetry groups and identified the general equations that are invariant under them. We also show that system (\ref{z1}) does not admit three parameter point symmetry groups, eventhough the subcase (\ref{intr2}) do admit such symmetries. We identify the following set of equations which admit two parameter symmetries, namely
\begin{eqnarray}
(i)\,\,\ddot{x}+f(x)\dot{x}^2+\left(\frac{\lambda_{2}\,(\lambda_{1}-2)}{\lambda_{1}}+g_{1}\,\left(\int{e^{\int{f\,dx}}\,dx+\lambda_3}\right)^{-\lambda_{1}}\right)\dot{x}+\frac{h_1\left(\int e^{\int f \, dx} \, dx+\lambda_3\right)^{1-2\lambda _1}}{e^{\int f \, dx}}\nonumber\\
-\frac{\lambda _2 g_1\left(\int e^{\int f \, dx} \, dx+\lambda_3\right)^{1-\lambda _1}}{\lambda _1e^{\int f \, dx}}-\frac{\lambda _2^2 \left(\lambda _1-1\right)\left(\int e^{\int f \, dx} \, dx+\lambda_3\right)}{\lambda _1^2e^{\int f \, dx}}=0,\label{intr4}
\end{eqnarray}
\begin{eqnarray}
(ii)\,\,\ddot{x}+f(x)\dot{x}^2+\left(g_1-2\lambda_2\ln{\left(\int e^{\int f(x) dx} \, dx+\lambda_3\right)}\right)\dot{x}+\frac{h_1\left(\int e^{\int f dx} \, dx+\lambda_3\right)}{e^{\int f dx}}\nonumber\\
+\frac{\lambda _2^2 \left(\int e^{\int f dx} \, dx+\lambda_3\right) \left(\ln\left(\int e^{\int f dx} \, dx+\lambda_3\right)\right)^2}{e^{\int f dx}}\nonumber\\
  -\frac{\lambda _2\left(\lambda _2+g_1\right)\left(\int e^{\int f dx} \, dx+\lambda_3\right) \ln \left(\int e^{\int f dx} \, dx+\lambda_3\right)}{e^{\int f dx}}=0,\label{intr5}
\end{eqnarray}
\begin{eqnarray}
(iii)\,\,\ddot{x}+f(x)\dot{x}^2+\left(g_1 e^{-\lambda _1\int e^{\int f(x) \, dx} \, dx}+\lambda _2\right)\dot{x}
+\left(h_1e^{-2\lambda_1\int{e^{\int{f(x)dx}}dx}}-\frac{g_1\lambda_2}{\lambda_1}e^{-\lambda_1\int{e^{\int{f(x)dx}}dx}}\right. \nonumber\\
\left.-\frac{\lambda_2^2}{\lambda_1}\right)e^{-\int f(x)dx}=0,\label{intr8}
\end{eqnarray}
where $g_1,h_1,\lambda_1,\,\lambda_2$ and $\lambda_3$ are constants. We also report the Lie point symmetries of these three equations. To prove the integrability of Eqs. (\ref{intr4})-(\ref{intr8}) we transform them into certain known integrable equations. It has already been proved that the transformed equations admit time independent conservative Hamiltonian description. In other words, these three equations which admit two parameter symmetries are Liouville integrable, after suitable transformations. Finally, the full equation (\ref{z1}) itself with arbitrary forms of $f(x),\,g(x)$ and $h(x)$ admits an obvious one parameter symmetry group, namely the time translational symmetry.

\subsection{Methodology}
In the course of identifying linearizable/integrable equations belonging to the general class of Eq. (\ref{z1}) through Lie point symmetry analysis, more importantly, we develop a systematic procedure to integrate the complicated system of determining equations in order to identify maximal/non-maximal Lie point symmetries. The procedure developed in the paper appears to be new to this kind of problems and applicable to a wider class of equations including the equations (\ref{intr1}) and (\ref{intr2}) which we have investigated earlier. We develop this procedure to overcome the obstacle in integrating the determining equations for the infinitesimals, $\xi$ and $\eta$. In the present study, it is very difficult to integrate the determining equations due to the presence of three arbitrary functions $f,\,g$ and $h$ involved in them. To confirm whether this system of equations will provide nontrivial solutions or not, to begin with, we investigate the integrability of this system of equations. By imposing the compatibility between the equations we end up at two equations both of them containing the infinitesimal symmetries $\xi$ and $\eta$ and their first derivatives only (which we call them as auxiliary equations). These two equations have some common coefficients. We then solve these two equations algebraically to obtain two integrability conditions for the determining equations, namely
\begin{eqnarray}
&& L_1=0,\,\,\,L_2=0,\,\,\,M,N\neq0,\label{z6.1}\\
&&M=0,\,\,\,N=0,\,\,\,L_1,L_2\neq0,\label{z6.2}
\end{eqnarray}
where the expressions for $L_1,L_2,M$ and $N$ are given below in Eqs. (\ref{z2}), (\ref{z3}), (\ref{z4}) and (\ref{z5}), respectively. From our analysis we conclude that one can integrate the determining equations under both these conditions.
 
We then analyze these two integrability conditions in detail. We observe the first integrability condition constitutes the linearizability criteria $(L_1=0,L_2=0)$ for Eq. (\ref{z1}) under invertible point transformation. Since any linearizable second order nonlinear ODEs admits maximal number of Lie point symmetries (which is again confirmed by the non-vanishing of the expression $M$ and $N$ that is $M,N\neq0$) we conclude that one can get maximal number of Lie point symmetries when we integrate the determining equations by imposing these conditions. From the expressions $L_1=0$ and $L_2=0$ we fix the explicit form of $g$ and $h$ in terms of $f$. The associated nonlinear ODE is given above in Eq. (\ref{z2.1}). 

We then move on to analyze the second integrability condition, namely $M=0,N=0$ and $L_1,L_2\neq0$. The conditions $L_1\neq0$ and $L_2\neq0$ implies that they admit lesser parameter symmetries only. By imposing a compatibility between the expressions $M=0$ and $N=0$ we arrive at an equation, $D_1(x)b+D_2(x)\dot{b}+D_3(x)\ddot{b}=0$ (see Eq. (\ref{ap13.1})), where the coefficients $D_1,D_2$ and $D_3$ are functions of $x$ alone and $b(t)$ is a symmetry function. After carefully examining the forms of the coefficients which appear in this equation we conclude that the left hand side can become zero only in the case $b=0$. Thus from the second integrability condition, we conclude that one of the arbitrary functions, $b$, appearing in the infinitesimals $\xi$ and $\eta$ vanishes and the overdetermined system of PDEs will provide only lesser number of Lie point symmetries. We expect that this procedure which we have developed in this paper can lead to a wider application.  
 
The plan of the paper is as follows. In Sec. II, we find the determining equations for (\ref{z1}). Using a set of rather general self consistency criteria, the general form for the maximal symmetry case is obtained in Sec. III and the corresponding symmetries are obtained. In Sec. IV, we discuss the integrability aspect of the general form of equation obtained in Sec. III. A specific example of physical interest for the linearizable case will be discussed in Sec. V. The non-maximal case with all its subcases will be discussed in Sec. VI. In Sec. VII, all the equations belonging to two parameter symmetry groups are obtained. The equivalence transformation for (\ref{z1}) is discussed in Sec. VIII. In Appendix  A, we apply the self consistency criteria and derive the relationship among the functions $f,\,g$ and $h$ whereas in Appendix B the first integrability condition is discussed. In Appendix C, we study the second integrability condition. Finally, we present our conclusion in Sec. IX.  


\section{Determining equations for the infinitesimal symmetries}
\label{sec2}
Let Eq. (\ref{z1}) be invariant under an one parameter group of symmetry transformations,
\begin{eqnarray}
&&T=t+\varepsilon \,\xi(t,x)+O(\epsilon^{2})\nonumber\\
&&X=x+\varepsilon \,\eta(t,x)+O(\epsilon^{2}),\quad \epsilon \ll 1,\label{a2}
\end{eqnarray}
where $\xi(t,x)$ and $\eta(t,x)$ are the infinitesimal point symmetry generators. An operator $G$ given by 
\begin{eqnarray}
G(t,x)=\xi(t,x)\,\frac{\partial }{\partial t}+\eta(t,x)\,\frac{\partial }{\partial x}\label{a3}
\end{eqnarray}
is said to be an infinitesimal generator of one parameter Lie point symmetry group of transformations for Eq. (\ref{z1}) iff
\begin{equation}
G^{(2)}(A)\mid_{A=0}=0,\label{a4}
\end{equation}
or equivalently
\begin{equation}
\Bigl(\xi\, \frac{\partial A}{\partial t}+\eta \,\frac{\partial A}{\partial x}+\eta^{(1)} \frac{\partial A}{\partial \dot{x}}+\eta^{(2)} \frac{\partial A}{\partial \ddot{x}}\Bigl)\,\mid_{A=0}=0,\label{a5}
\end{equation}
where
\begin{eqnarray}
G^{(2)}=G^{(1)}+\eta^{(2)}\frac{\partial}{\partial \ddot{x}},\quad G^{(1)}=G+\eta^{(1)}\frac{\partial}{\partial \dot{x}},\nonumber
\end{eqnarray}
\begin{eqnarray}
\eta^{(2)}=\frac{d\eta^{(1)}}{dt}-\ddot{x}\,\frac{d\xi}{dt},\quad \eta^{(1)}=\frac{d\eta}{dt}-\dot{x}\,\frac{d\xi}{dt}\label{a6}
\end{eqnarray}
and
\begin{equation}
\frac{d}{dt}=\frac{\partial}{\partial t}+\dot{x}\,\frac{\partial}{\partial x}.\label{a7}
\end{equation}

Substituting Eqs. (\ref{a6}) and (\ref{a7}) in Eq. (\ref{a5}) and equating different powers of $\dot{x}^m,\,m=0,\,1,\,2,\,3,$ to zero, we obtain the following set of determining equations,
\begin{eqnarray}
&&\xi_{xx}-f\xi_{x}=0,\label{a8} \\
&&\eta_{xx}+f\eta_{x}+\eta f_{x}-2\,\xi_{tx}+2g\xi_{x}=0, \label{a9}\\
&&2\,\eta_{tx}+\eta \,g_{x}+2\,f\,\eta_{t}+g\,\xi_{t}-\xi_{tt}+3\,h\,\xi_{x}=0, \label{a10}\\
&&\eta_{tt}+\eta\,h_{x}-h\, \eta_{x}+g\,\eta_{t}+2h\,\xi_{t}=0,\label{a11}
\end{eqnarray}
where subscripts denote partial derivatives. Now, the four equations, (\ref{a8})-(\ref{a11}), for the two unknown functions $\xi(t,x)$ and $\eta(t,x)$ constitute an overdetermined system of equations. Even then it is difficult to integrate these determining equations and obtain the explicit form of these infinitesimals $\xi$ and $\eta$. To integrate these complicated overdetermined system of PDEs we deduce the integrability conditions by imposing the compatibility between the various Eqs. (\ref{a8})-(\ref{a11}). Consequently it turns out that one can integrate the system of Eqs. (\ref{a8})-(\ref{a11}) under the following two conditions only, namely (fuller details are given in the Appendices A-C) 
 
\begin{eqnarray}
&&(i)\,\,L_1=0,\,\,\,L_2=0,\,\,\,M,N\neq0,\label{con1}\\
\text{and}\, 
&&(ii)\,\,L_1\neq0,\,\,\,L_2\neq0,\,\,\,M=0,\,\,\,N=0,\label{con2}
\end{eqnarray}
where the functions $L_1,\,L_2,\,M$ and $N$ are given by 
\begin{eqnarray}
&&L_1=g_{xx}-f g_x,\label{z2}\\ 
&&L_2=3 h_{xx}+3 h f_x+3 f h_x-2 g g_x,\label{z3}\\ 
&&M=\left(\frac{P}{\eta}\right)_{tt}\left(\frac{\xi_x}{\eta}\right)_t-\left(\frac{\xi_x}{\eta}\right)_{tt}\left(\frac{P}{\eta}\right)_t,\label{z4}\\
\text{and}
&&N=\left(\frac{\eta_t}{\eta}\right)_{tt}\left(\frac{Q}{\eta}\right)_t-\left(\frac{Q}{\eta}\right)_{tt}\left(\frac{\eta_t}{\eta}\right)_t.\label{z5} 
\end{eqnarray}
Here $P=2\eta_x+\xi_t$ and $Q=\eta_x+2\xi_t$. In Appendix A, the conditions (\ref{con1}) and (\ref{con2}) are deduced systematically. In Appendices B and C, the consequences of conditions (\ref{con1}) and (\ref{con2}) are respectively investigated. 

After examining carefully the above two integrability conditions, we conclude that in the first condition one can get maximal Lie point symmetries (since the conditions $L_1=0$ and $L_2=0$ are nothing but the necessary and sufficient condition for the linearization of (\ref{z1}) under point transformations as shown in the next section and see also Appendix B) and in the second condition ($M=0,\,N=0$) one has essentially non-maximal symmetries (since one of the infinitesimal symmetry functions becomes zero here) only. On the basis of these results we divide our analysis into 
two cases, $(i)\,L_1=0$ and $L_2=0$ $(M,\,N\neq0)$ and $(ii)\,L_1\neq0$ and $L_2\neq0$ $(M=0,\,N=0)$. We consider each one of the cases separately, integrate the determining equations, report the explicit form of the invariant equations and their infinitesimal symmetries. 

\section{Case i\, $\,L_1=0$ and $L_2=0$ - Eight Parameter Symmetries}
To start with we consider the case $L_1=0$ and $L_2=0$, with $M,\,N\neq0$, where $L_1,\,L_2,\,M$ and $N$ are given in Eqs. (\ref{z2})-(\ref{z5}), respectively. Before integrating the determining Eqs. (\ref{a8})-(\ref{a11}), we explore the form of $g$ and $h$ by solving the conditions $L_1=0$ and $L_2=0$. Since we have two equations with three unknowns, $f,\,g$ and $h$  in these two expressions, we fix $f$ as an arbitrary function and obtain $g$ and $h$ in terms of $f$. This in turn identifies the general ODE (\ref{z1}) which contains only one arbitrary function, say $f(x)$. We then consider this equation and integrate the associated determining equations (\ref{a8})-(\ref{a11}) and explore the infinitesimal symmetries $\xi$ and $\eta$ of it. 

\subsection{The General Equation}
In this subsection we display the explicit form of the ODE (\ref{z1}) that admits eight Lie point symmetries. 

To begin with, we recall the first condition $L_1=0$, that is $g_{xx}-fg_x=0,$ from which we obtain
\begin{eqnarray}
g=k_1\Im +k_2,\,\,\,\Im(x)=\int{F(x)dx},\,\,\,F(x)=e^{\int{f(x)dx}},\label{sym01}
\end{eqnarray}
where $k_1$ and $k_2$ are integration constants. Substituting the explicit form of $g$ in the second linearizability condition, namely 
$L_2=0$, we obtain an ODE for the other unknown function $h(x)$, that is
\begin{eqnarray}
h_{xx}+hf_x+fh_x=\frac{2}{3}gg_x=\frac{2}{3}k_1F(k_1\Im +k_2).\label{sym02}
\end{eqnarray}
Integrating (\ref{sym02}) once, we get
\begin{eqnarray}
h_x+hf=\frac{2}{3}k_1^2{\int{\Im\,F\,dx}}+\frac{2}{3}k_1k_2\int{Fdx}+k_3,\label{sym03}
\end{eqnarray}
where $k_3$ is another integration constant. 

To proceed further we rewrite the above equation in the form
\begin{eqnarray}
(hF)_x=\frac{1}{3}k_1^2\Im^2F+\frac{2}{3}k_1k_2\Im F+k_3F,\label{sym05}
\end{eqnarray}
which in turn provides the explicit form of $h$ by a straightforward integration. The explicit form of $h$ is given by
\begin{eqnarray}
h=\frac{k_1^2}{9}\,\frac{\Im^3}{F}+\frac{k_1k_2}{3}\,\frac{\Im^2}{F}+k_3\,\frac{\Im}{F}+\,\frac{k_4}{F},\label{sym06}
\end{eqnarray}
where $k_4$ is a constant of integration. 

Finally, rewriting Eqs. (\ref{sym01}) and (\ref{sym06}) in terms of $f$ and then substituting them in Eq. (\ref{z1}), we obtain the specific form of (\ref{z1}) which satisfies both the conditions (\ref{con1}) and (\ref{con2}) as
\begin{eqnarray}
\ddot{x}+f(x){\dot{x}}^2+\left(k_1\int{e^{\int{f(x)dx}}dx} +k_2\right)\dot{x}+e^{-\int{f(x)dx}}\left[\frac{k_1^2}{9}\left(\int{e^{\int{f(x)dx}}dx}\right)^3 \right. \nonumber\\
\left. +\frac{k_1k_2}{3}\left(\int{e^{\int{f(x)dx}}dx}\right)^2+k_3\,\left(\int{e^{\int{f(x)dx}}dx}\right)+k_4\right]=0.\label{sym07}
\end{eqnarray}
We will now prove that Eq. (\ref{sym07}) is invariant under the maximal number of Lie point symmetries.


\subsection{Eight Parameter Symmetries}
In this sub-section, we prove that Eq. (\ref{sym07}) is invariant under eight parameter Lie point symmetries. To do so we integrate the Eqs. (\ref{a8})-(\ref{a11}) with the previously determined forms of $g$ and $h$. By integrating Eq. (\ref{a8}), $\xi$ can be obtained unambiguously in terms of $f$, that is
\begin{eqnarray}
\xi=b(t)\Im(x)+a(t),\label{sym1}
\end{eqnarray}
where $\Im(x)=\int{F(x)dx}$ and $F(x)=e^{\int{f(x)dx}}$. Here $a(t)$ and $b(t)$ are arbitrary functions of $t$. 

Substituting (\ref{sym1}) and (\ref{sym01}) in (\ref{a9}) and integrating it twice, we get
\begin{eqnarray}
\eta=-\frac{{k_1b}}{3}\,\frac{\Im(x)^3}{F(x)}+(\dot{b}-k_2b)\,\frac{\Im(x)^2}{F(x)}+c\,\frac{\Im(x)}{F(x)}+\,\frac{d}{F(x)},\label{sym8}
\end{eqnarray}
where $b(t),\,c(t)$ and $d(t)$ are arbitrary functions of $t$. Substituting the forms of $g,h$ and $\eta$, vide Eqs. (\ref{sym01}), (\ref{sym06}) and (\ref{sym8}), respectively, in Eq. (\ref{a10}) and then equating the various functions of $f$ to zero we get the following two relations,
\begin{eqnarray}
&&3\ddot{b}-3k_2\dot{b}+3k_3b+k_1(\dot{a}+c)=0,\label{sym012}\\
&&2\dot{c}-\ddot{a}+k_2\dot{a}+k_1d+3k_4b=0.\label{sym013}
\end{eqnarray}
Again, substituting the forms of $g,h$ and $\eta$ from Eqs. (\ref{sym01}), (\ref{sym06}) and (\ref{sym8}) in the last determining Eq. (\ref{a11}) and then equating the various functions of $f$ to zero we get the following additional relations for the arbitrary functions $ a,\,b,\,c$ and $d$, namely
\begin{eqnarray}
&&\ddot{d}+k_2\dot{d}+k_3d-k_4(c-2\dot{a})=0,\label{sym014}\\
&&\ddot{c}+k_2\dot{c}+k_1\dot{d}+\frac{2}{3}k_1k_2d+2k_3\dot{a}+2k_2k_4b=0,\label{sym015}\\
&&\dddot{b}+k_1\dot{c}+\frac{k_1^2}{3}d+\frac{k_1k_2}{3}(c+2\dot{a})+(k_3-{k_2}^2)\dot{b}+(k_3k_2+k_1k_4)b=0.\label{sym016}
\end{eqnarray}
Solving (\ref{sym012})-(\ref{sym016}) one can obtain the explicit forms of the arbitrary function $ a,\,b,\,c$ and $d$ which all constitute the infinitesimals $\xi$ and $\eta$. 

We solve the determining equations in the following way. From (\ref{sym012}) and (\ref{sym016}) we get
\begin{eqnarray}
c&=&-\frac{3}{k_1}\ddot{b}+\frac{3}{k_1}k_2\dot{b}-\frac{3}{k_1}k_3b-\dot{a},\label{sym017}\\
d&=&\frac{6}{k_1^2}\dddot{b}-\frac{6}{k_1^2}k_2\ddot{b}+\frac{6}{k_1^2}k_3\dot{b}+\frac{3}{k_1}\ddot{a}-\frac{3k_4}{k_1}b-\frac{k_2}{k_1}\dot{a}.\label{sym018}
\end{eqnarray}
Now substituting the above forms of $c(t)$ and $d(t)$ and their derivatives into (\ref{sym014}) and (\ref{sym015}) and simplifying them we arrive at two coupled ODEs which involve only the functions $a$ and $b$, that is
\begin{eqnarray}
\frac{3}{k_1}\frac{d^4b}{dt^4}-\frac{2k_2}{k_1}\frac{d^3b}{dt^3}+(\frac{3k_3}{k_1}-\frac{k_2^2}{k_1})\frac{d^2b}{dt^2}+(\frac{k_2k_3}{k_1}-3k_4)\frac{db}{dt}+2\frac{d^3a}{dt^3}+(2k_3-\frac{2k_2^2}{3})\frac{da}{dt} = 0,\label{sym019}\\
\frac{6}{k_1^2}\frac{d^5b}{dt^5}+(\frac{12k_3}{k_1^2}-\frac{6k_2^2}{k_1^2})\frac{d^3b}{dt^3}+(\frac{6k_3^2}{k_1^2}-\frac{6k_4k_2}{k_1})\frac{db}{dt}+\frac{3}{k_1}\frac{d^4a}{dt^4}+\frac{2}{k_1}k_2\frac{d^3a}{dt^3}+(\frac{3k_3}{k_1}-\frac{k_2^2}{k_1})\frac{d^2a}{dt^2}\nonumber\\
+(3k_4-\frac{k_2k_3}{k_1})\frac{da}{dt}= 0.\label{sym020}
\end{eqnarray}
Using (\ref{sym019}) we can express certain derivatives of $a$ interms of $b$, that is
\begin{eqnarray}
\frac{d^3a}{dt^3}+(k_3-\frac{k_2^2}{3})\frac{da}{dt}=-\frac{3}{2k_1}\frac{d^4b}{dt^4}+\frac{k_2}{k_1}\frac{d^3b}{dt^3}+(\frac{k_2^2}{2k_1}-\frac{3k_3}{2k})\frac{d^2b}{dt^2}+(\frac{3k_4}{2}-\frac{k_2k_3}{2k_1})\frac{db}{dt}.\label{sym021}
\end{eqnarray}
Substituting (\ref{sym021}) into (\ref{sym020}) and rearranging  the latter we get
\begin{eqnarray}
\frac{da}{dt} = -\frac{1}{N_1}\Bigl(\frac{3}{2}\frac{d^5b}{dt^5}+\Bigl(\frac{15k_3}{2}-\frac{5k_2^2}{2}\Bigl)\frac{d^3b}{dt^3}+\Bigl(\frac{9k_1k_4}{2}-\frac{9k_2k_3}{2}+k_2^3\Bigl)\frac{d^2b}{dt^2} \nonumber\\
\qquad\qquad\qquad\qquad\qquad+\big(6k_3^2-3k_1k_2k_4-k_2^2k_3\big)\frac{db}{dt}\Bigl),\label{sym022}
\end{eqnarray}
where
\begin{equation}
N_1=3{k_1}^2k_4-3k_3k_1k_2+\frac{2}{3}k_1k_2^3.\label{sym023}
\end{equation}
Integrating (\ref{sym022}) once we get 
\begin{eqnarray}
a(t)=a_{1}-\frac{1}{N_1}\Bigl(\frac{3}{2}\frac{d^4b}{dt^4}+\biggl(\frac{15k_3}{2}-\frac{5k_2^2}{2}\biggl)\frac{d^2b}{dt^2}+\biggl(\frac{9}{2}k_1k_4-\frac{9}{2}k_2k_3+k_2^3\biggl)\frac{db}{dt}\nonumber\\
\qquad \qquad +(6k_3^2-3k_1k_2k_4-k_2^2k_3)b\Bigl),\label{sym024}
\end{eqnarray}
where $a_1$ is an integration constant.  

Substituting (\ref{sym022}) and its derivatives into (\ref{sym019}) and simplifying the latter we obtain the following seventh order ODE for the variable $b(t)$, namely
\begin{eqnarray}
\frac{d^7b}{dt^7}+\big(6k_3-2k_2^2\big)\frac{d^5b}{dt^5}+\big(9k_3^2-6k_3k_2^2+k_2^4\big)\frac{d^3b}{dt^3}+\big(3k_1^2k_4^2-6k_1k_2k_3k_4+\frac{4}{3}k_1k_2^3k_4\nonumber\\
+4k_3^3-k_3^2k_2^2\big)\frac{db}{dt}=0.\label{sym025} 
\end{eqnarray}
Introducing $\frac{db}{dt} = P$, where $P$ is the new dependent variable, (\ref{sym025}) can be rewritten as
\begin{eqnarray}
\frac{d^6P}{dt^6}+\big(6k_3-2k_2^2\big)\frac{d^4P}{dt^4}+\big(9k_3^2-6k_3k_2^2+k_2^4\big)\frac{d^2P}{dt^2}+\big(3k_1^2k_4^2-6k_1k_2k_3k_4+\frac{4}{3}k_1k_2^3k_4\nonumber\\
+4k_3^3-k_3^2k_2^2\big)P = 0. \label{sym026}
\end{eqnarray}
Equation (\ref{sym026}) is a sixth order linear ODE with constant coefficients whose solution can be found in the following way. Since (\ref{sym026}) contains constant coefficients, a general solution of this equation can be written in the form
\begin{equation}
P = b_2e^{m_1t}+b_3e^{m_2t}+b_4e^{m_3t}+b_5e^{m_4t}+b_6e^{m_5t}+b_7e^{m_6t},\label{sym027}
\end{equation}
where ${b_i}^{'}s,~i=2,\ldots,7$, are integration constants and ${m_{i}}^{'}s,~i=1,\ldots,6$, are roots of the characteristic equation
\begin{eqnarray}
m^6+(6k_3-2k_2^2)m^4+(9k_3^2-6k_3k_2^2+k_2^4)m^2+(3k_1^2k_4^2-6k_1k_2k_3k_4+\frac{4}{3}k_1k_2^3k_4\nonumber\\
+4k_3^3-k_3^2k_2^2) = 0.\label{sym028}
\end{eqnarray}
Now choosing $m^2=\chi$, we can rewrite the sixth power polynomial equation (\ref{sym028}) as a cubic polynomial equation in $\chi$, namely, 
\begin{equation}
\chi^3+a\chi^2+b\chi+c= 0,\label{sym029}
\end{equation}
where 
\begin{eqnarray}
&&a=6k_3-2k_2^2,\quad  b=9k_3^2-6k_3k_2^2+k_2^4,\nonumber\\
&&c=3k_1^2k_4^2-6k_1k_2k_3k_4+\frac{4}{3}k_1k_2^3k_4+4k_3^3-k_3^2k_2^2.\label{sym030}
\end{eqnarray}
Introducing a transformation
\begin{equation}
\chi=Y-\frac{a}{3}\label{sym031}
\end{equation}
 in (\ref{sym029}), the later equation can be brought to the standard form
\begin{equation}
Y^3+pY+q = 0,\label{sym032}
\end{equation}
where the new constants, $p$ and $q$ are related to the old constants, $a,b$ and $c$, by the following relations 
\begin{equation}
p=b-\frac{a^2}{3}, \quad q = \frac{2}{27}a^3-\frac{ab}{3}+c.\label{sym033}
\end{equation}
The cubic equation (\ref{sym032}) has one real root and two complex conjugate roots, namely 
\begin{eqnarray}
\hspace{-.2in}Y_1 = A+B, \quad Y_2 = -\frac{(A+B)}{2}+i\frac{\sqrt{3}(A-B)}{2},\quad Y_3 = -\frac{(A+B)}{2}-i\frac{\sqrt{3}(A-B)}{2},\label{sym034}
\end{eqnarray}
where 
\begin{eqnarray} 
A = \left(-\frac{q}{2}+\sqrt{Q}\right)^{\frac{1}{3}}, \quad B = \left(-\frac{q}{2}-\sqrt{Q}\right)^{\frac{1}{3}}, \quad Q = \left(\frac{p}{3}\right)^3+\left(\frac{q}{2}\right)^2\label{sym035}
\end{eqnarray}
so that the roots of (\ref{sym029}) can now be expressed using the relation (\ref{sym031}), that is, 
\begin{eqnarray}
&&\chi_1 = A+B-\frac{a}{3}, \quad \chi_2 = -\frac{(A+B)}{2}-\frac{a}{3}+i\frac{\sqrt{3}(A-B)}{2}, \nonumber\\
&& \chi_3 = -\frac{(A+B)}{2}-\frac{a}{3}-i\frac{\sqrt{3}(A-B)}{2}.\label{sym036}
\end{eqnarray}
From the identity $m^2=\chi$ we are in a position to write down the six roots of (\ref{sym028}), that is,
\begin{eqnarray}
&&\hat{m}_1^2 =  \chi_1 = A+B-\frac{a}{3}, \quad \hat{m}_2^2 = \chi_2 = C+iD, \quad \hat{m}_3^2 = \chi_3 = C-iD, \label{sym037}
\end{eqnarray}
where again for simplicity we introduced the constants $C$ and $D$ which can be fixed from the relation  
\begin{equation}
C = -\frac{(A+B)}{2}-\frac{a}{3}, \quad D = \frac{\sqrt{3}(A-B)}{2}.\label{sym038}
\end{equation}
Equation (\ref{sym037}) furnishes the six roots, 
\begin{eqnarray}
 &&m_{1}=+\sqrt{A+B-\frac{a}{3}},\quad m_{2}  =-\sqrt{A+B-\frac{a}{3}},\nonumber\\
 &&m_{3}=\sqrt{C^2+D^2}\left[\cos\phi+i\sin\phi\right],\quad m_{4}=-\sqrt{C^2+D^2}\left[\cos\phi+i\sin\phi\right],\nonumber\\
 &&m_{5}=\sqrt{C^2+D^2}\left[\cos\phi-i\sin\phi\right],\quad m_{6}=-\sqrt{C^2+D^2}\left[\cos\phi-i\sin\phi\right],\nonumber
\end{eqnarray}
where
\begin{equation}
\phi = \left(\frac{tan^{-1}(D/C)}{2}\right).\label{sym039}
\end{equation}
With the help of these ${m_{i}}^{'}s,\; i=1,...,6$, a general solution for (\ref{sym026}) can be written in the form
\begin{eqnarray}
P=&b_2e^{\alpha_1 t}+b_3e^{-\alpha_1 t}+e^{\alpha_2 t}(b_4\cos\alpha_3t+b_5\sin\alpha_3t)+e^{-\alpha_2t}(b_6\cos\alpha_3t-b_7\sin\alpha_3t),\label{sym040}
\end{eqnarray}
where ${b_i}^{'}s,\; i=2,\ldots,7$, are integration constants and $\alpha_{1}=m_{1}$, whereas $\alpha_2,\alpha_3$ are the real and imaginary parts of $m_3$, namely $\alpha_2=\sqrt{C^2+D^2}\cos\phi,\;\;\alpha_3=\sqrt{C^2+D^2}\sin\phi$.

From the identity $\frac{db}{dt} = P$ the function $b(t)$ can be deduced easily by integrating it, that is, as
\begin{eqnarray}
b(t)&=&b_1+\frac{b_2e^{\alpha_1 t}}{\alpha_1}-\frac{b_3e^{-\alpha_1t}}{\alpha_1}+\frac{b_4e^{\alpha_2t}(\alpha_2\cos\alpha_3t
+\alpha_3\sin\alpha_3 t)}{(\alpha_2^2+\alpha_3^2)}+ \frac{b_5e^{\alpha_2t}(\alpha_2\sin\alpha_3t-\alpha_3\cos\alpha_3 t)}{(\alpha_2^2+\alpha_3^2)}\nonumber\\
&&+\frac{b_6e^{-\alpha_2t}(\alpha_3\sin\alpha_3 t-\alpha_2\cos\alpha_3t)}{(\alpha_2^2+\alpha_3^2)}+\frac{b_7e^{-\alpha_2t}(\alpha_2\sin\alpha_3t+\alpha_3\cos\alpha_3 t)}{(\alpha_2^2+\alpha_3^2)},\label{sym041}
\end{eqnarray}
where $b_1$ is an integration constant.  Once $b(t)$ is known the function $a(t)$ can be fixed from the relation (\ref{sym024}) as
\begin{eqnarray}
a(t)&=&a_1-N_2b_1+\beta_1b_2e^{\alpha_1 t}+\beta_2b_3e^{-\alpha_1t}+b_4e^{\alpha_2t}(\beta_3\cos\alpha_3t+\beta_4\sin\alpha_3 t)\nonumber\\
&&+b_5e^{\alpha_2t}(\beta_3\sin\alpha_3t-\beta_4\cos\alpha_3t)+b_6e^{-\alpha_2t}(\beta_5\cos\alpha_3t+\beta_4\sin\alpha_3 t)\nonumber\\
&&+b_7e^{-\alpha_2t}(\beta_4\cos\alpha_3t-\beta_5\sin\alpha_3 t),\label{sym042}
\end{eqnarray}
where $a_1$ is an integration constant and $N_2=(6k_3^2-3k_1k_2k_4-k_2^2k_3)/N_1$. For simplicity in the expression $a(t)$ we have introduced the constants $\beta_i,\;i=1,2,\ldots,5$.  These new constants are related to the old parameters through the following relations
\begin{eqnarray}
\beta_1&=&-\frac{1}{N_1}\Bigl(k_2^3+\frac{3}{2}\alpha_1^3-\frac{5}{2}k_2^2\alpha_1-\frac{9}{2}(k_2k_3-k_1k_4)+\frac{15}{2}\alpha_1k_3 
+\frac{6}{\alpha_1}k_3^2-\frac{3}{\alpha_1}k_1k_2k_4-\frac{k_2^2}{\alpha_1}k_3\Bigl),\nonumber\\
\beta_2&=&-\frac{1}{N_1}\Bigl(k_2^3-\frac{3}{2}\alpha_1^3+\frac{5}{2}k_2^2\alpha_1-\frac{9}{2}(k_2k_3-k_1k_4)-\frac{15}{2}\alpha_1k_3 -\frac{6}{\alpha_1}k_3^2+\frac{3}{\alpha_1}k_1k_2k_4+\frac{k_2^2}{\alpha_1}k_3\Bigl),\nonumber\\
\beta_3&=&-\frac{1}{N_1}\Bigl(k_2^3+\frac{3}{2}\alpha_2^3-\frac{5}{2}k_2^2\alpha_2-\frac{9}{2}(k_2k_3-k_1k_4)-\frac{9}{2}\alpha_2\alpha_3^2+\frac{15}{2}\alpha_2k_3-\frac{1}{(\alpha_2^2+\alpha_3^2)}(k_2^2\alpha_2k_3\nonumber\\
&&-6\alpha_2k_3^2+3k_1k_2\alpha_2k_4)\Bigl),\nonumber\\
\beta_4&=&-\frac{1}{N_1}\Bigl(\frac{3}{2}\alpha_3^3+\frac{5}{2}k_2^2\alpha_3-\frac{9}{2}\alpha_2^2\alpha_3-\frac{15}{2}\alpha_3k_3 -\frac{(k_2^2\alpha_3k_3-6\alpha_3k_3^2+3k_1k_2\alpha_3k_4)}{(\alpha_2^2+\alpha_3^2)}\Bigl),\nonumber\\
\beta_5&=&-\frac{1}{N_1}\Bigl(k_2^3-\frac{3}{2}\alpha_2^3+\frac{5}{2}k_2^2\alpha_2-\frac{9}{2}(k_2k_3-k_1k_4)+\frac{9}{2}\alpha_2\alpha_3^2 -\frac{15}{2}\alpha_2k_3+\frac{1}{(\alpha_2^2+\alpha_3^2)}(k_2^2\alpha_2k_3\nonumber\\
&&-6\alpha_2k_3^2+3k_1k_2\alpha_2k_4)\Bigl).\label{sym043}
\end{eqnarray}
With the above forms of $a(t)$ and $b(t)$, the function $c(t)$ can be deduced using the relation (\ref{sym017}) as 
\begin{eqnarray}
c(t)&=&-\frac{3k_3}{k_1}b_1+\gamma_1b_2e^{\alpha_1t}+\gamma_2b_3e^{-\alpha_1t}+ b_4e^{\alpha_2t}(\gamma_3\cos\alpha_3t+\gamma_4\sin\alpha_3 t)\nonumber\\
&&+ b_5e^{\alpha_2t}(\gamma_3\sin\alpha_3t-\gamma_4\cos\alpha_3t)+b_6e^{-\alpha_2t}(\gamma_5\cos\alpha_3t+\gamma_6\sin\alpha_3 t)\nonumber\\
&&+b_7e^{-\alpha_2t}(\gamma_6\cos\alpha_3t-\gamma_{5}\sin\alpha_3t),\label{sym044}  
\end{eqnarray}
where again we have defined new constants ${\gamma_i}^{'}s,\; i=1,2,\ldots,5$, just for simplicity as
\begin{eqnarray}
\gamma_1&=&-\left(\frac{3\alpha_1}{k_1}-\frac{3k_2}{k_1}+\frac{3k_3}{k_1\alpha_1}+\alpha_1\beta_1\right),\quad \gamma_2 = \left(\frac{3\alpha_1}{k_1}+\frac{3k_2}{k_1}+\frac{3k_3}{k_1\alpha_1}+\alpha_1\beta_2\right), \nonumber\\
\gamma_3&=&-\left(\frac{3\alpha_2}{k_1}-\frac{3k_2}{k_1}+\frac{3k_3\alpha_2}{k_1(\alpha_2^2+\alpha_3^2)}
+\alpha_2\beta_3+\alpha_3\beta_4\right),\nonumber\\
\gamma_4&=&\left(\frac{3\alpha_3}{k_1}-\frac{3k_3\alpha_3}{k_1(\alpha_2^2+\alpha_3^2)}+\alpha_3\beta_3-\alpha_2\beta_4\right),\nonumber\\
\gamma_5&=&\left(\frac{3\alpha_2}{k_1}+\frac{3k_2}{k_1}+\frac{3k_3\alpha_2}{k_1(\alpha_2^2+\alpha_3^2)}-\alpha_3\beta_4+\alpha_2\beta_5\right),\nonumber\\ 
\gamma_6&=&\left(\frac{3\alpha_3}{k_1}-\frac{3k_3\alpha_3}{k_1(\alpha_2^2+\alpha_3^2)}+\alpha_2\beta_4+\alpha_3\beta_5\right).\label{sym045}
\end{eqnarray}
Finally the function $d(t)$ can be derived from the relation (\ref{sym018}) by simply substituting the forms of $a(t)$ and $b(t)$ and their derivatives into it, that is,  
\begin{eqnarray}
d(t)&=&-\frac{3k_4}{k_1}b_1+\delta_1b_2e^{\alpha_1t}+\delta_2b_3e^{-\alpha_1t}+ b_4e^{\alpha_2t}(\delta_3\cos\alpha_3t+\delta_4\sin\alpha_3 t)\nonumber\\
&&+b_5e^{\alpha_2t}(\delta_3\sin\alpha_3t-\delta_4\cos\alpha_3t)+b_6e^{-\alpha_2t}(\delta_5\cos\alpha_3t+\delta_6\sin\alpha_3 t)\nonumber\\ 
&&+b_7e^{-\alpha_2t}(\delta_6\cos\alpha_3t-\delta_{5}\sin\alpha_3 t).\label{sym046}
\end{eqnarray}
Again for simplicity we have defined the constants ${\delta_i}'s,\;i=1,2,\ldots,6$, which can be related to the older parameters through the following relations,
\begin{eqnarray}
\delta_1&=&-\left(\frac{6\alpha_1}{k_1^2}(k_2-\alpha_1)+\frac{\alpha_1\beta_1}{k_1}(k_2-3\alpha_1)+\frac{3}{k_1}\biggl(\frac{k_4}{\alpha_1}-\frac{2k_3}{k_1}\biggl)\right),\nonumber\\
\delta_2&=&\left(\frac{6\alpha_1}{k_1^2}(k_2+\alpha_1)+\frac{\alpha_1\beta_2}{k_1}(k_2+3\alpha_1)+\frac{3}{k_1}\biggl(\frac{k_4}{\alpha_1}+\frac{2k_3}{k_1}\biggl)\right),\nonumber\\
\delta_3&=&\frac{1}{k_1}\Biggl(-k_2\alpha_2(\beta_3+\frac{6}{k_1})+3(\alpha_2^2-\alpha_3^2)(\beta_3+\frac{2}{k_1})-\alpha_3\beta_4(k_2-6\alpha_2)-\frac{3\alpha_2k_4}{(\alpha_2^2+\alpha_3^2)}+\frac{6k_3}{k_1}\Biggl),\nonumber\\
\delta_4&=&\frac{\alpha_3}{k_1}\Biggl(k_2(\beta_3+\frac{6}{k_1})-6\alpha_2(\beta_3+\frac{2}{k_1})-\frac{\alpha_2\beta_4}{\alpha_3}(k_2-3\alpha_2)-3\alpha_3\beta_4-\frac{3k_4}{(\alpha_2^2+\alpha_3^2)}\Biggl),\nonumber\\
\delta_5&=&\frac{1}{k_1}\Biggl(k_2\alpha_2(\beta_5+\frac{6}{k_1})+3(\alpha_2^2-\alpha_3^2)(\beta_5+\frac{2}{k_1})-\alpha_3\beta_4(k_2+6\alpha_2)+\frac{3\alpha_2k_4}{(\alpha_2^2+\alpha_3^2)}+\frac{6k_3}{k_1}\Biggl),\nonumber\\
\delta_6&=&\frac{\alpha_3}{k_1}\left(k_2(\beta_5+\frac{6}{k_1})+6\alpha_2(\beta_5+\frac{2}{k_1})+\frac{\alpha_2\beta_4}{\alpha_3}(k_2+3\alpha_2)\right.
-3\alpha_3\beta_4-\frac{3k_4}{(\alpha_2^2+\alpha_3^2)}\biggl).\label{sym047}
\end{eqnarray}
Finally, substituting the forms of $a,b,c$ and $d$ into (\ref{sym1}) and (\ref{sym8}) we obtain the infinitesimal symmetries associated with 
Eq. (\ref{sym07}) as 
\begin{eqnarray}
\xi&=&a_1-N_2b_1+\beta_1b_2e^{\alpha_1 t}+\beta_2b_3e^{-\alpha_1t}+b_4e^{\alpha_2t}(\beta_3\cos\alpha_3t+\beta_4\sin\alpha_3 t)\nonumber\\
&&+b_5e^{\alpha_2t}(\beta_3\sin\alpha_3 t-\beta_4\cos\alpha_3t)+b_6e^{-\alpha_2t}(\beta_5\cos\alpha_3t+\beta_4\sin\alpha_3 t)\nonumber\\
&&+b_7e^{-\alpha_2t}(\beta_4cos\alpha_3t-\beta_{5}\sin\alpha_3 t)+\Im(x)\,\Bigl(b_1+\frac{b_2}{\alpha_1}e^{\alpha_1t}-\frac{b_3}{\alpha_1}e^{-\alpha_1t}\nonumber\\
&&+\frac{b_4e^{\alpha_2t}}{(\alpha_2^2+\alpha_3^2)}(\alpha_2\cos\alpha_3t+\alpha_3\sin\alpha_3t)+\frac{b_5e^{\alpha_2t}}{(\alpha_2^2+\alpha_3^2)}(\alpha_2\sin\alpha_3t-\alpha_3\cos\alpha_3 t)\nonumber\\
&&+\frac{b_6e^{-\alpha_2t}}{(\alpha_2^2+\alpha_3^2)}(\alpha_3\sin\alpha_3 t-\alpha_2\cos\alpha_3t)+\frac{b_7e^{-\alpha_2t}}{(\alpha_2^2+\alpha_3^2)}(\alpha_2\sin\alpha_3t+\alpha_3\cos\alpha_3 t)\Bigl),\nonumber\\
\eta&=&-\frac{k_1}{3}\Biggl(b_1+\frac{b_2}{\alpha_1}e^{\alpha_1t}-\frac{b_3}{\alpha_1}e^{-\alpha_1t}+\frac{b_4e^{\alpha_2t}}{(\alpha_2^2+\alpha_3^2)}(\alpha_2\cos\alpha_3t+\alpha_3\sin\alpha_3 t)\nonumber\\
&&+\frac{b_5e^{\alpha_2t}}{(\alpha_2^2+\alpha_3^2)}(\alpha_2\sin\alpha_3t-\alpha_3\cos\alpha_3 t)+\frac{b_6e^{-\alpha_2t}}{(\alpha_2^2+\alpha_3^2)}(\alpha_3\sin\alpha_3t-\alpha_2\cos\alpha_3t)\nonumber\\
&&+\frac{b_7e^{-\alpha_2t}}{(\alpha_2^2+\alpha_3^2)}(\alpha_2\sin\alpha_3 t+\alpha_3\cos\alpha_3t)\Biggl)\frac{\Im(x)^3}{F}+\Biggl(b_2e^{\alpha_1 t}+b_3e^{-\alpha_1t}\nonumber\\
&&+b_4e^{\alpha_2 t}\cos\alpha_3t+b_5e^{\alpha_2 t}\sin\alpha_3t+b_6e^{-\alpha_2 t}\cos\alpha_3t+b_7e^{-\alpha_2t}\sin\alpha_3t\Biggl)\frac{\Im(x)^2}{F}\nonumber\\
&&+\biggl(-\frac{3k_3}{k_1}b_1+\gamma_1b_2e^{\alpha_1t}+\gamma_2b_3e^{-\alpha_1t}+b_4e^{\alpha_2t}(\gamma_3\cos\alpha_3t+\gamma_4\sin\alpha_3 t)+b_5e^{\alpha_2t}(\gamma_3\sin\alpha_3t\nonumber\\
&&-\gamma_4\cos\alpha_3t)+ b_6e^{-\alpha_2t}(\gamma_5\cos\alpha_3t+\gamma_6\sin\alpha_3 t)+b_7e^{-\alpha_2t}(\gamma_6\cos\alpha_3t-\gamma_{5}\sin\alpha_3t)\biggl)\frac{\Im(x)}{F}\nonumber
\end{eqnarray}
\begin{eqnarray}
&&+\biggl(-\frac{3k_4}{k_1}b_1+\delta_1b_2e^{\alpha_1 t}+\delta_2b_3e^{-\alpha_1t}+b_4e^{\alpha_2t}(\delta_3\cos\alpha_3t+\delta_4\sin\alpha_3 t)\nonumber\\
&&+ b_5e^{\alpha_2t}(\delta_3\sin\alpha_3t-\delta_4\cos\alpha_3t)+b_6e^{-\alpha_2t}(\delta_5\cos\alpha_3t+\delta_6\sin\alpha_3 t)\nonumber\\
&&+b_7e^{-\alpha_2t}(\delta_6\cos\alpha_3t-\delta_{5}\sin\alpha_3 t)\biggl)\frac{1}{F}.\label{sym048}
\end{eqnarray}
We note that in the above $a_1,b_i,~ i=2,...,7$, are the eight integration constants arising while integrating the determining equations. These eight
integration constants, as usual, fix the eight infinitesimal generators of the form
\begin{eqnarray}
X_1&=&\frac{\partial}{\partial t},\nonumber\\
X_2&=&(\Im-N_2)\frac{\partial}{\partial t}+\left(-\frac{3k_4}{k_1F}-\frac{3k_3\Im(x)}{k_1F}-\frac{k_1\Im(x)^3}{3F}\right)\frac{\partial}{\partial x},\nonumber\\
X_3&=&e^{\alpha_1t}\left[\left(\beta_1+\frac{\Im}{\alpha_1}\right)\frac{\partial}{\partial t}+\left(\frac{\delta_1}{F}+\gamma_1\frac{\Im(x)}{F}+\frac{\Im(x)^2}{F}-\frac{k_1}{3\alpha_1}\frac{\Im(x)^3}{F}\right)\frac{\partial}{\partial x}\right],\nonumber\\
X_4&=&e^{-\alpha_1t}\left[\left(\beta_2-\frac{\Im}{\alpha_1}\right)\frac{\partial}{\partial t}+\left(\frac{\delta_2}{F}+\gamma_2\frac{\Im(x)}{F}+\frac{\Im(x)^2}{F}+\frac{k_1}{3\alpha_1}\frac{\Im(x)^3}{F}\right)\frac{\partial}{\partial x}\right],\nonumber\\
X_5&=&e^{\alpha_2 t}\biggr[(\beta_3\cos\alpha_3 t+\beta_4\sin\alpha_3t+\frac{1}{(\alpha_2^2+\alpha_3^3)}(\alpha_2\cos\alpha_3t+\alpha_3\sin\alpha_3t)\Im)\frac{\partial}{\partial t}\nonumber\\
&&+\Bigl(\frac{\delta_3\cos\alpha_3t+\delta_4\sin\alpha_3t}{F}+(\gamma_3\cos\alpha_3t+\gamma_4\sin\alpha_3t)\frac{\Im(x)}{F}+\cos\alpha_3t\frac{\Im(x)^2}{F}\nonumber\\
&&-\frac{k_1}{3(\alpha_2^2+\alpha_3^2)}(\alpha_2\cos\alpha_3t+\alpha_3\sin\alpha_3t)\frac{\Im(x)^3}{F}\Bigl)\frac{\partial}{\partial x}\biggr],\nonumber
\\
X_6&=&e^{\alpha_2t}\biggr[(\beta_3\sin\alpha_3t-\beta_4\cos\alpha_3t+\frac{1}{(\alpha_2^2+\alpha_3^3)}(\alpha_2\cos\alpha_3t-\alpha_3\sin\alpha_3t)\Im)\frac{\partial}{\partial t}\nonumber\\
&&+\Bigl(\frac{\delta_3\sin\alpha_3t-\delta_4\cos\alpha_3t}{F}+(\gamma_3\sin\alpha_3t-\gamma_4\cos\alpha_3t)\frac{\Im(x)}{F}+\sin\alpha_3t\frac{\Im(x)^2}{F}\nonumber\\
&&-\frac{k_1}{3(\alpha_2^2+\alpha_3^2)}(\alpha_2\cos\alpha_3t-\alpha_3\sin\alpha_3t)\frac{\Im(x)^3}{F}\Bigl)\frac{\partial}{\partial x}\biggr],\nonumber\\
X_7&=&e^{-\alpha_2t}\biggr[(\beta_5\cos\alpha_3t+\beta_4\sin\alpha_3t+\frac{1}{(\alpha_2^2+\alpha_3^3)}(\alpha_3\sin\alpha_3t-\alpha_2\cos\alpha_3t)\Im)\frac{\partial}{\partial t}\nonumber\\
&&+\Bigl(\frac{\delta_5\cos\alpha_3t+\delta_6\sin\alpha_3t}{F}+(\gamma_5\cos\alpha_3t+\gamma_6\sin\alpha_3t)\frac{\Im(x)}{F}+\cos\alpha_3t\frac{\Im(x)^2}{F}\nonumber\\
&&-\frac{k_1}{3(\alpha_2^2+\alpha_3^2)}(\alpha_3\sin\alpha_3t-\alpha_2\cos\alpha_3t)\frac{\Im(x)^3}{F}\Bigl)\frac{\partial}{\partial x}\biggr],\nonumber\\
X_8&=&e^{-\alpha_2t}\biggr[(\beta_4\cos\alpha_3t-\beta_{5}\sin\alpha_3t+\frac{1}{(\alpha_2^2+\alpha_3^3)}(\alpha_2\sin\alpha_3t+\alpha_3\cos\alpha_3t)\Im)\frac{\partial}{\partial t}\nonumber\\
&& +\Bigl(\frac{\delta_6\cos\alpha_3t-\delta_{5}\sin\alpha_3t}{F}+(\gamma_6\cos\alpha_3t-\gamma_{5}\sin\alpha_3t)\frac{\Im(x)}{F}+\sin\alpha_3t\frac{\Im(x)^2}{F}\nonumber\\
&&-\frac{k_1}{3(\alpha_2^2+\alpha_3^2)}(\alpha_2\sin\alpha_3t+\alpha_3\cos\alpha_3t)\frac{\Im(x)^3}{F}\Bigl)\frac{\partial}{\partial x}\biggr].\label{sym049}
\end{eqnarray}
The generators $X_1,X_2,.....,X_8$ form an $sl(3,R)$ algebra.


\section{Integrability and Linearization of Eq. (\ref{sym07})}
Here, in this section we study the integrability properties of Eq. (\ref{sym07}) admitting eight Lie point symmetries, with the help of the so called modified Prelle-Singer method \cite{vkc1} applicable to second order ODEs. This method helps to derive first integrals from which the general solution  can be deduced and to establish complete integrability, which has been discussed in detail by Chandrasekar \textit{et al} \cite{vkc1}. 

\subsection{Integrals of motion and general solution}
Let us rewrite Eq. (\ref{sym07}) in the form 
\begin{eqnarray}
\ddot{x}=-f{\dot{x}}^2-(k_1\Im+k_2)\dot{x}-\frac{k_1^2}{9}\frac{\Im^3}{F}-\frac{k_1k_2}{3}\frac{\Im^2}{F}-k_3\frac{\Im}{F}-\frac{k_4}{F}=\phi(x,\dot{x}),\label{int1}
\end{eqnarray}
where $\Im=\int{fdx}$ and $F=e^{\int{fdx}}$. If Eq. (\ref{int1}) admits a first integral $I(t,x,\dot{x})=C$, where $C$ is a constant on the solutions, then the total differential can be written as
\begin{eqnarray}
dI=I_tdt+I_xdx+I_{\dot{x}}d\dot{x}=0.\label{int2}
\end{eqnarray}
Now, Eq. (\ref{int1}) can be rewritten as
\begin{eqnarray}
(\phi+S\dot{x})dt-Sdx-d\dot{x}=0,\label{int3}
\end{eqnarray}
where we have added a null term $S(t,x,\dot{x})\dot{x}dt-S(t,x,\dot{x})dx$ to (\ref{int1}). Hence on the solutions (\ref{int2}) and (\ref{int3}) must be proportional. Multiplying (\ref{int3}) by the factor $R(t,x,\dot{x})$ that acts as the integrating factor for (\ref{int3}), we have on the solutions,
\begin{equation}
dI=R(\phi+S\dot{x})dt-RSdx-Rd\dot{x}=0.\label{int4}
\end{equation}
Comparing (\ref{int2}) with (\ref{int4}), on the solutions, we have the following relations
\begin{eqnarray}
I_{t}=R(\phi+\dot{x}S),\quad I_{x}=-RS,\quad I_{\dot{x}}=-R. \label{int5}
\end{eqnarray}
Applying the compatibility conditions $I_{tx}=I_{xt},I_{t\dot{x}}=I_{\dot{x}t}, I_{x\dot{x}}=I_{\dot{x}x}$ between (\ref{int5}), we get the determining equations for the null form $S$ and the integrating factor $R$ as
\begin{eqnarray}
S_t+\dot{x}S_x-\left[f{\dot{x}}^2+(k_1\Im+k_2)\dot{x}+\frac{k_1^2}{9}\frac{\Im^3}{F}+\frac{k_1k_2}{3}\frac{\Im^2}{F}+k_3\frac{\Im}{F}+\frac{k_4}{F}\right]S_{\dot{x}}=f_x{\dot{x}}^2+k_1F\dot{x}\nonumber\\
+\frac{k_1^2}{9}(3\Im^2-f\frac{\Im^3}{F})+\frac{k_1k_2}{3}(2\Im-f\frac{\Im^2}{F})+k_3(1-f\frac{\Im}{F})+f\frac{k_4}{F}\nonumber\\
-\left(2f\dot{x}+k_1\Im+k_2\right)S+S^2,\label{int6}
\end{eqnarray}
\begin{align}
\hspace{-.3in}R_t+\dot{x}R_x-\left[f{\dot{x}}^2+(k_1\Im+k_2)\dot{x}+\frac{k_1^2}{9}\frac{\Im^3}{F}+\frac{k_1k_2}{3}\frac{\Im^2}{F}+k_3\frac{\Im}{F}+\frac{k_4}{F}\right]R_{\dot{x}}=(2f\dot{x}+k_1\Im\nonumber\\
+k_2-S)R,\label{int7}\\
\hspace{-.3in}R_x=SR_{\dot{x}}-RS_{\dot{x}}.\label{int8}
\end{align}
Solving Eqs. (\ref{int6})-(\ref{int8}) in the same way as was done by Chandrasekar \textit{et al} \cite{vkc1}, one can obtain two independent sets of expressions for $S$ and $R$ in the forms
\begin{eqnarray}
&&S_1=\frac{k_1\left(\int{e^{\int{fdx}}dx}\right)+3\alpha}{3}-\frac{k_1\dot{x}e^{\int{fdx}}}{k_1\left(\int{e^{\int{fdx}}dx}\right)+3\alpha}+f\dot{x}, \nonumber\\
&&R_1=\frac{C_0e^{\int{fdx}}(k_1\left(\int{e^{\int{fdx}}dx}\right)+3\alpha)e^{\mp\hat{\alpha}t}}{(3k_1\dot{x}e^{\int{fdx}}-\frac{\hat{\beta}\pm\hat{\alpha}}{2}(3k_1\left(\int{e^{\int{fdx}}dx}\right)+9\alpha)+(k_1\left(\int{e^{\int{fdx}}dx}\right)+3\alpha)^2)^2}, 
\end{eqnarray}
and \begin{eqnarray}
&&S_2=\frac{k_1\left(\int{e^{\int{fdx}}dx}\right)+3\beta}{3}-\frac{k_1\dot{x}e^{\int{fdx}}}{k_1\left(\int{e^{\int{fdx}}dx}\right)+3\beta}+f\dot{x},\nonumber\\
&&R_2=\frac{C_0e^{\int{fdx}}(k_1\left(\int{e^{\int{fdx}}dx}\right)+3\beta)e^{\frac{\hat{\beta}\pm\hat{\alpha}}{2}t}}{3k_1\dot{x}e^{\int{fdx}}-\frac{\hat{\beta}\mp\hat{\alpha}}{2}(3k_1\left(\int{e^{\int{fdx}}dx}\right)+9\alpha)+(k_1\left(\int{e^{\int{fdx}}dx}\right)+3\alpha)^2},
\end{eqnarray}  
where $C_0=9k_1\hat{\alpha},\,\,\,\alpha^3-k_2\alpha^2+\alpha k_3-\frac{k_1k_4}{3}=0,\,\,\hat{\alpha}=\sqrt{-3\alpha^2+2\alpha k_2+k_2^2-4k_3},\,\,\hat{\beta}=3\alpha-k_2$ and $\beta= \frac{-\alpha+k_2\pm\hat{\alpha}}{2}$. Now, the integrals of motion for (\ref{z1}) can be obtained from the expression 
\begin{align}
I(t,x,\dot{x}) & = \int R(\phi+\dot{x}S)dt   -\int \left( RS+\frac{d}{dx}\int R(\phi+\dot{x}S)dt \right) dx  \notag\\
  & -\int \left\{R+\frac{d}{d\dot{x}} \left[\int R (\phi+\dot{x}S)dt- \int \left(RS+\frac{d}{dx}\int R(\phi+\dot{x}S)dt\right)dx\right]\right\}d\dot{x},
  \label{met13}
\end{align}
as can be deduced by solving (\ref{int5}). Then the two time dependent integrals associated with (\ref{sym07}) or (\ref{int1}) turn out to be
\begin{eqnarray}
I_1 &=&e^{\mp\hat{\alpha}t}\left(\frac{3k_1e^{\int{fdx}}\dot{x}-\frac{\hat{\beta}\mp\hat{\alpha}}{2}(3k_1\int{e^{\int{fdx}}dx}+9\alpha)+(k_1\int{e^{\int{fdx}}dx}+3\alpha)^2}{3k_1e^{\int{fdx}}\dot{x}-\frac{\hat{\beta}\pm\hat{\alpha}}{2}(3k_1\int{e^{\int{fdx}}dx}+9\alpha)+(k_1\int{e^{\int{fdx}}dx}+3\alpha)^2}\right),\label{int31}\\
I_2 &=&{\left(\frac{3k_1e^{\int{fdx}}\dot{x}-3k_1\int{e^{\int{fdx}}dx}(\alpha-k_2)+k_1^2\left(\int{e^{\int{fdx}}dx}\right)^2+9\alpha^2-9\alpha k_2+9k_3}{3k_1e^{\int{fdx}}\dot{x}-\frac{\hat{\beta}\mp\hat{\alpha}}{2}(3k_1\int{e^{\int{fdx}}dx}+9\alpha)+(k_1\int{e^{\int{fdx}}dx}+3\alpha)^2}\right)}\nonumber\\
&&\times \frac{-2\hat{\alpha}e^{\frac{\hat{\beta}\pm\hat{\alpha}}{2}t}}{\hat{\beta}\pm\hat{\alpha}}.\label{int32}
\end{eqnarray}
Note that from out of the two time dependent integrals (\ref{int31}) and (\ref{int32}) one can also obtain a time dependent integral $I$ as $I=I_1^2I_2$, from which a time independent Hamiltonian can also be constructed. 

Finally the solution of (\ref{z1}) can be written from (\ref{int31}) and (\ref{int32}) as
\begin{eqnarray}
\int{e^{\int{fdx}}dx}=-\frac{3\alpha}{k_1}+\frac{1}{k_1}{\left(\frac{6(3\alpha^2-2\alpha k_2+k_3)(1-I_1e^{\pm\hat{\alpha}t})}{\hat{\beta}(1-I_1e^{\pm\hat{\alpha}t})\pm (\hat{\beta}\pm\hat{\alpha})I_1I_2e^{\frac{-\hat{\beta}\pm\hat{\alpha}}{2}t}\pm \hat{\alpha}{(1+I_1e^{\pm\hat{\alpha}t})}}\right)}.\label{solf}
\end{eqnarray}
From (\ref{solf}) it is clear that if the form of the arbitrary function $f$ is given then the general solution can in principle be obtained by carrying out the integral on the left hand side of (\ref{solf}).


\subsection{Linearizing transformation}
While it has already been proved that Eq. (\ref{sym07}) admits eight symmetry generators and so is expected to be linearizable 
(see Appendix B), we need to identify the exact linearizing transformation for Eq. (\ref{sym07}). We derive the linearizing 
transformation from the first integral. For the sake of completeness we briefly discuss the method of deriving linearizing 
transformation from first the integrals \cite{vkc1}. 

Let us consider that there exists a first integral $I=F(t,x,\dot{x})$ for Eq. (\ref{sym07}). Then we can rewrite this first integral 
by splitting the function $F$ into a product of two functions as
\begin{eqnarray}
I=F\left(\frac{1}{G_2(t,x,\dot{x})}\,\frac{d}{dt}G_1(t,x)\right).\label{line1}
\end{eqnarray}
From the above equation one can identify $G_1(t,x)$ as the new dependent variable and the integral of $G_2(t,x,\dot{x})$ over time as the new independent variable. Then we can write
\begin{eqnarray}
w=G_1(t,x),\qquad z=\int_0^{t}{G_2(t',x,\dot{x})dt'}.\label{line2}
\end{eqnarray} 
Correspondingly Eq. (\ref{line1}) can be written as
\begin{eqnarray}
\hat{I}=\frac{dt}{dz}\,\frac{dw}{dt}=\frac{dw}{dz},\label{line3}
\end{eqnarray}
where $\hat{I}_1$ as a constant. Hence, we have
\begin{eqnarray}
\frac{d^2w}{dz^2}=0,\label{line4}
\end{eqnarray}
which is nothing but free particle equation. Hence the new variables $w$ and $z$ provide a linearizing transformation that helps in transforming the given second order nonlinear ODE into a linear second order ODE.

Following the above procedure we can rewrite the linearizing transformation for (\ref{sym07}) by rewriting the first integral given by (\ref{int31}) as
\begin{eqnarray}
I_1=\frac{-e^{\frac{-\hat{\beta}\mp\hat{\alpha}}{2}t}(k_1\int{e^{\int{fdx}}dx}+3\alpha)^2}{\left(3k_1e^{\int{fdx}}\dot{x}-\frac{\hat{\beta}\pm\hat{\alpha}}{2}(3k_1\int{e^{\int{fdx}}dx}+9\alpha)+(k_1\int{e^{\int{fdx}}dx}+3\alpha)^2\right)}\nonumber\\
\times\left[\frac{d}{dt}\left(\left(\frac{-3}{k_1\int{e^{\int{fdx}}dx}+3\alpha}+\frac{\hat{\beta}\pm\hat{\alpha}}{2(3\alpha^2-2\alpha k_2+k_3)}\right)e^{\frac{\hat{\beta}\mp\hat{\alpha}}{2}t}\right)\right].\label{line5}
\end{eqnarray}
 The linearizing point transformation for (\ref{sym07}) will be then
\begin{eqnarray}
&&w = \left(\frac{-3}{k_1\int{e^{\int{fdx}}dx}+3\alpha}+\frac{\hat{\beta}\pm\hat{\alpha}}{2(3\alpha^2-2\alpha k_2+k_3)}\right)e^{\frac{\hat{\beta}\mp\hat{\alpha}}{2}t},\nonumber\\
&&z =\left(\frac{-3}{k_1\int{e^{\int{fdx}}dx}+3\alpha}+\frac{\hat{\beta}\mp\hat{\alpha}}{2(3\alpha^2-2\alpha k_2+k_3)}\right)
e^{\frac{\hat{\beta}\pm\hat{\alpha}}{2}t},
\end{eqnarray}
so that $\frac{d^2w}{dz^2}=0$, which is indeed the free particle equation. Thus (\ref{sym07}) or (\ref{int1}) stands linearized. To our knowledge, the linearizing transformations to (\ref{sym07}) are being reported for the first time in the literature.


\section{Example of a linearizable equation}
In this section we consider an example belonging to Eq. (\ref{sym07}) by fixing the form of $f(x)$. Let us consider the case 
$f=\frac{\lambda}{x}$, where $\lambda$ is a parameter. In this case the exact form of Eq. (\ref{sym07}) turns out that 
\begin{eqnarray}
\ddot{x}+\frac{\lambda}{x}\, {\dot{x}}^2+\left(\frac{k_1x^{\lambda+1}}{\lambda+1}+k_2\right)\dot{x}
+\frac{k_1^2x^{2\lambda +3}}{9(\lambda+1)^3}+\frac{k_1k_2x^{\lambda+2}}{3(\lambda+1)^2}+\frac{k_3x}{\lambda+1}+k_4x^{-\lambda }=0.\label{le1}
\end{eqnarray}
As the determining equations will be the same as given by Eq. (\ref{sym012})-(\ref{sym016}), the forms of the functions 
$a(t), b(t),c(t)$ and $d(t)$ will also be the same. Therefore the infinitesimal symmetries associated with (\ref{le1}) 
can be obtained directly from Eq. (\ref{sym048}), which lead to the eight infinitesimal generators. On the other hand 
following the procedure as discussed in the previous section one can obtain the integrals of motion and hence the 
exact solution for Eq. (\ref{le1}) as 
\begin{eqnarray}
x(t)=\left[(\lambda+1)\left(-\frac{3\alpha}{k_1}+\frac{1}{k_1}{\bigg(\frac{6(3\alpha^2-2\alpha k_2+k_3)(1-I_1e^{\pm\hat{\alpha}t})}{\hat{\beta}(1-I_1e^{\pm\hat{\alpha}t})\pm (\hat{\beta}\pm\hat{\alpha})I_1I_2e^{\frac{-\hat{\beta}\pm\hat{\alpha}}{2}t}\pm \hat{\alpha}{(1+I_1e^{\pm\hat{\alpha}t})}}\bigg)}\right)\right]^{\frac{1}{\lambda+1}}.\label{solfex}
\end{eqnarray}


\section{Non-Maximal Symmetries: Case $b=0$}
Next we consider the second integrability condition $L_1\neq0,\,L_2\neq0,\,M=0,\,N=0$, given by Eq. (\ref{con2}). As proved in Appendix C, this case corresponds to the condition that the symmetry function $b(t)=0$. To start with we consider the general case $f\neq0,\,g\neq0$ and $h\neq0$. Substituting $b=0$ in Eq. (\ref{sym1}), we get 
\begin{eqnarray}
\xi(t)=a(t).\label{nm01}
\end{eqnarray}
Using this in Eq. (\ref{a9}) and integrating it twice, we get the form of $\eta $ as
\begin{eqnarray}
\eta=c G_2+ d G_3,\label{nm1}
\end{eqnarray}
where $G_2=\frac{\Im(x)}{F(x)},\,G_3=\frac{1}{F(x)},\,\Im=\int{Fdx}$ and $F=e^{\int{fdx}}$ and $c(t)$ and $d(t)$ are arbitrary functions of $t$. Now, with the help of Eqs. (\ref{nm01}) and (\ref{nm1}), Eqs. (\ref{a10}) and (\ref{a11}) can be rewritten as 
\begin{eqnarray}
2\,(\dot{c}\,G_{2x}+\dot{d}\,G_{3x})+g_{x}\,(c\,G_{2}+d\,G_{3})+2\,f\,(\dot{c}\,G_{2}+\dot{d}\,G_{3})-\ddot{a}+g\,\dot{a}=0,\label{b1}
\end{eqnarray}
and
\begin{eqnarray}
\ddot{c}\,G_{2}+\ddot{d}\,G_{3}+h_{x}\,(c\,G_{2}+d\,G_{3})-h\,(c\,G_{2x}+d\,G_{3x})+g\,(\dot{c}\,G_{2}+\dot{d}\,G_{3})+2\,h\,\dot{a}=0,\label{b2}
\end{eqnarray}
respectively. It is to be noted that Eq. (\ref{b1}) involves the function $f(x)$. So for any arbitrary form of the function $f$, one can obtain the form of the function $g$ by solving Eq. (\ref{b1}). Now, substituting the obtained form of $g$ into Eq. (\ref{b2}) and integrating the underlying equation we can get the expression of $h$. Hence we see that fixing the form of $f$ alone is enough to fix the form of Eq. (\ref{z1}) in this case.

However, it has already been proved that a second order ODE admits only one, two, three, or eight parameter Lie point symmetries \cite{ibra,maho}. The eight parameter symmetry has already been discussed in the previous section, where all the four functions $a(t),\,b(t),\,c(t)$ and $d(t)$ are nonzero. Now for the non-maximal case with $b = 0$, we explicitly show in the following that only one and two parameter symmetries exist when all the three functions $f\neq0,\,g\neq0$ and $h\neq0$ and obtain their specific forms.

For any arbitrary form of the functions $f,\,g$ and $h$ the simplest solution for Eqs. (\ref{b1}) and (\ref{b2}) is $a=a_0=$ constant and the other symmetry parameters $b,\,c,\,d=0$. It means that, one gets the time translation operator $X=\frac{\partial}{\partial t}$ for arbitrary form of the functions. Hence the general Eq. (\ref{z1}) with three arbitrary functions, $f,\,g$ and $h$, is invariant under the one parameter Lie point symmetry group, as expected. 

Now, with $c,\,d\neq0$, Eq. (\ref{b1}) can be rewritten as
\begin{eqnarray}
g_{x}+g\,\Bigl(\frac{\dot{a}}{c\,G_{2}+d\,G_{3}}\Bigl)+\frac{2\,f\,(\dot{c}\,G_{2}+\dot{d}\,G_{3})+2\,(\dot{c}\,G_{2x}+\dot{d}\,G_{3x})-\ddot{a}}{c\,G_{2}+d\,G_{3}}=0.\label{c1}
\end{eqnarray}
Since $g(x)$ is a function of $x$ alone, we define
\begin{eqnarray}
\frac{\dot{a}}{c}=\lambda_{1},\,\,\,\frac{\dot{c}}{c}=\lambda_{2},\,\,\,\frac{d}{c}=\lambda_{3},\qquad c\neq0,\label{c2}
\end{eqnarray}
where $\lambda_1,\,\lambda_2$ and $\lambda_3$ are arbitrary constants. It is to be noted that dividing by $c$ or $d$ in (\ref{c2}) does not correspond to any different set of equations as one can redefine the parameters and can return back to the original set of equations.

Then Eq. (\ref{c1}) can be rewritten as
\begin{eqnarray}
g_{x}+g\,\Bigl(\frac{\lambda_{1}}{G_{2}+\lambda_{3}\,G_{3}}\Bigl)+\lambda_2\,\Bigl(\frac{2\,(G_{2x}+\lambda_{3}\,G_{3x})-\lambda_{1}}{G_{2}+\lambda_{3}\,G_{3}}+2\,f\Bigl)=0,\label{c3}
\end{eqnarray}
where we have used the relations $\dot{a}=\lambda_1c,\,\ddot{a}=\lambda_1\dot{c}=\lambda_1\lambda_2c$ and $\dot{d}=\lambda_3\dot{c}=\lambda_2\lambda_3c$. Now, rewriting Eq. (\ref{c3}) in terms of $f$ after using the forms of $G_2$ and $G_3$, we get
\begin{eqnarray}
g_{x}+g\,\Bigl(\frac{\lambda_{1}e^{\int{f\,dx}}}{\int{e^{\int{f\,dx}}\,dx}+\lambda_{3}}\Bigl)+\frac{\lambda_{2}\,(2-\lambda_{1})\,e^{\int{f\,dx}}}{\int{e^{\int{f\,dx}}\,dx}+\lambda_{3}}=0.\label{c4}
\end{eqnarray}
Integrating the above equation once we obtain
\begin{eqnarray}
g=\frac{\lambda_{2}\,(\lambda_{1}-2)}{\lambda_{1}}+g_{1}\,\Bigl(\int{e^{\int{f\,dx}}\,dx}+\lambda_{3}\Bigl)^{-\lambda_{1}},\label{c5}
\end{eqnarray}
where $g_{1}$ is an integration constant. It is to be noted that $g_1,\,\lambda_{1},\,\lambda_{2}$ and $\lambda_{3}$ are now system parameters as they are contributing to the form of $g$ in (\ref{z1}).

Next, from Eq. (\ref{c2}) we can write the explicit form of the functions $a(t),\,c(t)$ and $d(t)$ for $\lambda_1\neq0,\,\lambda_2\neq0$ and $\lambda_3\neq0$ as
\begin{eqnarray}
a=a_1+\frac{\lambda_1}{\lambda_2}\,c_1e^{\lambda_2t},\,\,\,c=c_1e^{\lambda_2t},\,\,\,d=\lambda_3c_1e^{\lambda_2t},\label{c02a}
\end{eqnarray}
where $a_1$ and $c_1$ are symmetry parameters. Since $\lambda_1,\,\lambda_2$ and $\lambda_3$ are system parameters appearing in the 
form of the function $g$ in (\ref{c5}), from the above forms of $a(t),\,c(t)$ and $d(t)$, we identify only two arbitrary parameters 
$a_1$ and $c_1$ as corresponding to a two parameter symmetry group. This implies that there exists no three parameter symmetry 
group for system (\ref{z1}). In the next section we will explore all the equations corresponding to two parameter symmetry group. 


\section{Two Parameter Symmetries}
Now, to deduce all the invariant equations belonging to (\ref{z1}) under two parameter symmetries we consider 
two possibilities $(i)\, c\neq0,\,a,\,d\neq0$ and $(ii)\,c=0,\,a,\,d\neq0$ and also the subcases in both of them. 
As noted above system (\ref{z1}) always admits translational symmetry and hence we do not take $a=0$ 
(from (\ref{sym1})). On the other hand if both $c$ and $d$ are simultaneously zero, then $\dot{a}=0$ from (\ref{b1}) 
and (\ref{b2}), which corresponds to one parameter symmetry group. Hence we need not consider the case $a\neq0,\,c,d=0$. 
Finally, the case $a,\,c\neq0,\,d=0$ is a subcase of case $(i)$. Hence we need to discuss only the cases $(i)$ and $(ii)$ along 
with their possible subcases.   


\subsection{Integrable Equation for $\bf{c,\,a,\,d\neq0}$}
It is clear from Eq. ({\ref{c2}}) that for the case $c,\,a,\,d\neq0$ we have eight different possibilities for the parameters 
$\lambda_1,\,\lambda_2$ and $\lambda_3$ as $(i)\,\lambda_1\neq0,\,\lambda_2\neq0,\,\lambda_3\neq0,\,
(ii)\,\lambda_1=0,\,\lambda_2\neq0,\,\lambda_3\neq0,\,(iii)\,\lambda_1\neq0,\,\lambda_2\neq0,\,\lambda_3=0,\,
(iv)\,\lambda_1\neq0,\,\lambda_2=0,\,\lambda_3\neq0,\,(v)\,\lambda_1\neq0,\,\lambda_2=0,\,\lambda_3=0,\,
(vi)\,\lambda_1=0,\,\lambda_2\neq0,\,\lambda_3=0,\,(vii)\,\lambda_1=0,\,\lambda_2=0,\,\lambda_3\neq0,\,
(viii)\,\lambda_1=0,\,\lambda_2=0,\,\lambda_3=0$. After analyzing these eight possibilities we find that separate study is needed only 
for two possibilities. These two cases are $(i)\,\lambda_1\neq0,\,\lambda_2\neq0,\,\lambda_3\neq0,\,
(ii)\,\lambda_1=0,\,\lambda_2\neq0,\,\lambda_3\neq0$. The rest of the cases are either subcases of these two cases or they belong to a 
subcase in the linearizable case, so there is no need to discuss these cases separately. Now, each of these two possibilities 
are discussed separately in the following.


\subsubsection*{$(i)\,\text{Case\,1}:\,\lambda_1\neq0,\,\lambda_2\neq0,\,\lambda_3\neq0$}
With $\lambda_1\neq0,\,\lambda_2\neq0,\,\lambda_3\neq0$, the form of the $g$ is given by Eq. (\ref{c5}). 
Now, solving (\ref{c2}) for $\lambda_1\neq0,\,\lambda_2\neq0,\,\lambda_3\neq0$, we get the explicit forms of the functions $a(t)$ and $c(t)$ as given by Eq. (\ref{c02a}) which leads to a two parameter symmetry group. Substituting the forms of $a,\,c$ and $d$ in (\ref{b2}), we arrive at the relation
\begin{eqnarray}
h_{x}+h\,\Bigl(\frac{2\,\lambda_{1}-1}{G_{2}+\lambda_3G_{3}}+f\Bigl)+\lambda_{2}\,(\lambda_{2}+g)=0.\label{d2}
\end{eqnarray}
Rewriting Eq. (\ref{d2}) in terms of $f$ we get
\begin{eqnarray}
\hspace{-.2in}h_{x}+\Biggl(\frac{(2\,\lambda_{1}-1)\,e^{\int{f\,dx}}}{\int{e^{\int{f\,dx}}\,dx}+\lambda_3}+f\Biggl)\,h+\lambda_{2}\,\Biggl(\frac{2\,\lambda_{2}\,(\lambda_{1}-1)}{\lambda_{1}}+g_{1}\,\Bigl(\int{e^{\int{f\,dx}}\,dx+\lambda_3}\Bigl)^{-\lambda_{1}}\Biggl)=0.\label{d3}
\end{eqnarray}
Integrating Eq. (\ref{d3}) we get the form of the function $h$ as
\begin{eqnarray}
h= \frac{h_1\left(\int e^{\int f \, dx} \, dx+\lambda_3\right)^{1-2\lambda _1}}{e^{\int f \, dx}}-\frac{\lambda _2 g_1\left(\int e^{\int f \, dx} \, dx+\lambda_3\right)^{1-\lambda _1}}{\lambda _1e^{\int f \, dx}}\nonumber\\
-\frac{\lambda _2^2 \left(\lambda _1-1\right)\left(\int e^{\int f \, dx} \, dx+\lambda_3\right)}{\lambda _1^2e^{\int f \, dx}},\label{d4}
\end{eqnarray}
where $h_1$ is an integration constant. With the help of the forms of $g$ and $h$ [vide Eqs. (\ref{c5}) and (\ref{d4})], Eq. (\ref{z1}) takes the form
\begin{eqnarray}
\hspace{-.2in}\ddot{x}+f(x)\dot{x}^2+\left(\frac{\lambda_{2}\,(\lambda_{1}-2)}{\lambda_{1}}+g_{1}\,\left(\int{e^{\int{f\,dx}}\,dx+\lambda_3}\right)^{-\lambda_{1}}\right)\dot{x}+\frac{h_1\left(\int e^{\int f \, dx} \, dx+\lambda_3\right)^{1-2\lambda _1}}{e^{\int f \, dx}}\nonumber\\
-\frac{\lambda _2 g_1\left(\int e^{\int f \, dx} \, dx+\lambda_3\right)^{1-\lambda _1}}{\lambda _1e^{\int f \, dx}}
-\frac{\lambda _2^2 \left(\lambda _1-1\right)\left(\int e^{\int f \, dx} \, dx+\lambda_3\right)}{\lambda _1^2e^{\int f \, dx}}=0.\label{d5}
\end{eqnarray}
Eq. (\ref{d5}) admits the following infinitesimal symmetries,
\begin{eqnarray}
\xi=a_{1}+\frac{\lambda_1}{\lambda_{2}}\,c_{1}\,e^{\lambda_{2}\,t},\,\,\,\eta=c_{1}\,e^{\lambda_{2}\,t-\int{fdx}}\left(\int e^{\int f\,dx}\,dx+\lambda_3\right).\label{d7}
\end{eqnarray}
The corresponding infinitesimal generators are
\begin{eqnarray}
X_1=\frac{\partial}{\partial t},\,\,\,X_2=e^{\lambda_2t}\left[\frac{\lambda_1}{\lambda_2}\,\frac{\partial}{\partial t}+e^{-\int{fdx}}\left(\int e^{\int f\,dx}\,dx+\lambda_3\right)\,\frac{\partial}{\partial x}\right].\label{d8}
\end{eqnarray}

To study the integrability aspect of Eq. (\ref{d5}) we introduce a transformation $X=\int{e^{\int{fdx}}dx}+\lambda_3$ in it. Then (\ref{d5}) can be rewritten as
\begin{eqnarray}
\ddot{X}+(k_2+g_1X^{q})\dot{X}+\frac{k_2^2(q+1)}{(q+2)^2}X+\frac{k_2g_1}{(q+2)}X^{q+1}+h_1X^{2q+1}=0,\label{d6.1}
\end{eqnarray}
where $\lambda_1=-q$ and $\lambda_2=\frac{k_2q}{q+2}$. Now, introducing a transformation
\begin{eqnarray}
w = Xe^{\frac{k_2}{(q+2)}t},\;\;\; z =-\frac{(q+2)}{qk_2}e^{-\frac{qk_2}{(q+2)}t}, \label{ps34aa}
\end{eqnarray}
Eq. (\ref{d6.1}) can be transformed to the form
\begin{eqnarray}
w''+\alpha w^qw'+\beta w^{2q+1}=0, \label{ps34b}
\end{eqnarray}
where $\alpha = f_1$ and $\beta =g_1$. Different aspects of Eq. (\ref{ps34b}) have been studied by many authors and has been 
proved to be integrable for certain parametric choices of $\beta$ \cite{lem,feix,vkc7}. Recently, it has been shown that for 
(\ref{ps34b}) a time independent Hamiltonian for all values of $\alpha$ and $\beta$ \cite{vkc7,vkc5} can be identified as
\begin{eqnarray}
H=\left\{\label{emden_ham}
\begin{array}{ll}
\frac{(r-1)}{(r-2)}p^{\frac{(r-2)}{(r-1)}}
-\frac{(r-1)}{r} \hat{\alpha} pw^{q+1}, & \alpha^2>4\beta (q+1)\\
\frac{\hat{\alpha}}{2}pw^{q+1}+\log(\frac{1}{p}),&\alpha^2=4\beta (q+1)\\
\frac{1}{2}\log\left[\frac{w^{2(q+1)}}{(q+1)^2}
\sec^2[\frac{\omega}{(q+1)}w^{q+1}p]\right]-\frac{\hat{\alpha}}{2}pw^{q+1},&\alpha^2<4\beta (q+1),
\end{array}
\right.
\label{A1}
\end{eqnarray}
\vskip 4pt
\noindent where the corresponding canonically conjugate momentum is defined by
\begin{eqnarray}
p=\left\{
\begin{array}{ll}
\left(\dot{w}+\frac{(r-1)}{r}
\hat{\alpha}w^{q+1}\right)^{(1-r)}, \qquad \qquad \qquad \qquad \quad \alpha^2 \ge 4\beta (q+1)\\
\frac{(q+1)}{\omega w^{q+1}}\tan^{-1}\left[\frac{\alpha w^{q+1}
+2(q+1)\dot{w}}{2\omega w^{q+1}}\right]  \qquad \qquad \qquad \quad \; \alpha^2 < 4\beta (q+1),
\end{array}
\right. \label{A2}
\end{eqnarray}
where $r = \frac{\alpha}{2\beta(q+1)}(\alpha\pm\sqrt{\alpha^2-4\beta(q+1)},\,\omega=\frac{1}{2}\sqrt{4\beta(q+1)-\alpha^2}$ and $\hat{\alpha} = \frac{\alpha}{q+1}$. For more details one can see Ref. 31. The time independent Hamiltonian ensures the Liouville integrability of (\ref{ps34b}) or (\ref{d6.1}) and so (\ref{d5}).

\subsubsection*{Example:}
Let us consider the function $f$ as $f=\frac{k_1}{x}$ and $\lambda_3=0$, then $g$ from Eq. (\ref{c5}) takes the form
\begin{eqnarray}
g=k_2+k_3x,\label{ex12}
\end{eqnarray}
where $-\lambda_1(k_1+1)=1,g_1=k_3$ and $\frac{\lambda_2(\lambda_1-1)}{\lambda_1}=k_2$. Substituting the form of $g$ in Eq. (\ref{d4}), we get
\begin{eqnarray}
h=k_4x^3+\frac{k_2k_3}{2k_1+3}x^2+\frac{k_2^2(k_1+2)}{(2k_1+3)^2}x,\label{ex13}
\end{eqnarray}
where $h_1$ is replaced by $k_4$ for convenience. Hence with these forms of $g$ and $h$, Eq. (\ref{z1}) takes the form
\begin{eqnarray}
\ddot{x}+\frac{k_1}{x}{\dot{x}}^2+(k_2+k_3x)\dot{x}+k_4x^3+\frac{k_2k_3}{2k_1+3}x^2+\frac{k_2^2(k_1+2)}{(2k_1+3)^2}x=0.\label{ex14}
\end{eqnarray}
The infinitesimal symmetries for (\ref{ex14}) are
\begin{eqnarray}
\xi=a_{1}+\frac{\lambda_1}{\lambda_{2}}\,c_{1}\,e^{\lambda_{2}\,t},\,\,\,\eta=c_{1}\,\frac{x}{k_1+1}\,e^{\lambda_{2}\,t}.\label{ex15}
\end{eqnarray}
The corresponding infinitesimal generators are given by
\begin{eqnarray}
X_1=\frac{\partial}{\partial t},\,\,\,X_2=e^{\lambda_2t}\left[\frac{\lambda_1}{\lambda_2}\,\frac{\partial}{\partial t}+\frac{x}{k_1+1}\,\frac{\partial}{\partial x}\right].\label{ex16}
\end{eqnarray}
Following the procedure as discussed in the previous section one can prove the Livoulle integrability of the above Eq. (\ref{ex14}) which has already been done in Ref. 32.

\subsubsection*{$(ii)\,\text{Case\,2}:\,\lambda_1=0,\,\lambda_2\neq0,\,\lambda_3\neq0$}
Earlier we considered $\lambda_1\neq0$ for deriving the form of the function $f$. Now let us consider the case $\lambda_1=0$. For this case the explicit forms of the functions $a(t),\,c(t)$ and $d(t)$ from (\ref{c2}) work out to be
\begin{eqnarray}
a=a_1,\,\,\,\,\,c=c_1e^{\lambda_2t},\,\,\,\,d=\lambda_3c_1e^{\lambda_2t},\label{e1}
\end{eqnarray}
where $a_1$ and $c_1$ are arbitrary parameters which lead to a two parameter symmetry group. Substituting the forms of $a$ and $c$ in Eq. (\ref{b1}) we obtain the form of the function $g$ as
\begin{eqnarray}
g=g_1-2\lambda_2\ln{\left(\int e^{\int f dx} \, dx+\lambda_3\right)},\label{e2}
\end{eqnarray}
where $g_1$ is a constant of integration. Now Eq. (\ref{b2}) can be written with the help of Eqs. (\ref{e1}) and (\ref{e2}) as
\begin{eqnarray}
h_x+h\left(f-e^{\int f dx}\left(\int{e^{\int f dx}\,dx}+\lambda_3\right)^{-1}\right)+\lambda_2(\lambda_2+g)=0.\label{e3}
\end{eqnarray} 
Integrating (\ref{e3}) after substituting the form of $g$ from (\ref{e2}) we arrive at
\begin{eqnarray}
h&=&\frac{h_1\left(\int e^{\int f dx} \, dx+\lambda_3\right)}{e^{\int f dx}}-\frac{\lambda _2\left(\lambda _2+g_1\right)\left(\int e^{\int f dx} \, dx+\lambda_3\right) \ln \left(\int e^{\int f dx} \, dx+\lambda_3\right)}{e^{\int f dx}}\nonumber\\
&&+\frac{\lambda _2^2 \left(\int e^{\int f dx} \, dx+\lambda_3\right) \ln\left(\int e^{\int f dx} \, dx+\lambda_3\right)^2}{e^{\int f dx}}.\label{e4}
\end{eqnarray}
where $h_1$ is an integration constant. Now, with the help of these forms of $g$ and $h$ [vide (\ref{e2}) and (\ref{e4})], Eq. (\ref{z1}) takes the form
\begin{eqnarray}
\ddot{x}+f(x)\dot{x}^2+\left(g_1-2\lambda_2\ln{\left(\int e^{\int f dx} \, dx+\lambda_3\right)}\right)\dot{x}+\frac{h_1\left(\int e^{\int f dx} \, dx+\lambda_3\right)}{e^{\int f dx}}\nonumber\\
 -\frac{\lambda _2\left(\lambda _2+g_1\right)\left(\int e^{\int f dx} \, dx+\lambda_3\right) \ln \left(\int e^{\int f dx} \, dx+\lambda_3\right)}{e^{\int f dx}}\nonumber\\
+\frac{\lambda _2^2 \left(\int e^{\int f dx} \, dx+\lambda_3\right) \left(\ln\left(\int e^{\int f dx} \, dx+\lambda_3\right)\right)^2}{e^{\int f dx}}
 =0.\label{e5}
\end{eqnarray}
 which is invariant under a two parameter Lie point symmetry group with infinitesimal symmetries,
\begin{eqnarray}
\xi=a_1,\,\,\,\eta=c_{1}\,e^{\lambda_{2}\,t-\int{fdx}}\left(\int e^{\int f\,dx}\,dx+\lambda_3\right).\label{e6}
\end{eqnarray}
The corresponding infinitesimal generators read as
\begin{eqnarray}
X_1=\frac{\partial}{\partial t},\,\,\,X_2=e^{\lambda_2\,t-\int{fdx}}\left(\int e^{\int f\,dx}\,dx+\lambda_3\right)\,\frac{\partial}{\partial x}.\label{e7}
\end{eqnarray}

Introducing the transformation $X=\int e^{\int f\,dx}\,dx+\lambda_3$ in (\ref{e5}), we get
\begin{eqnarray}
\ddot{X}+(g_1-2\lambda_2\ln{X})\dot{X}+h_1X-(\lambda_2^2+\lambda_2g_1)X\ln{X}+\lambda_2^2X(\ln{X})^2=0.\label{e8}
\end{eqnarray}
The integrability of Eq. (\ref{e8}) can be proved in the following way by following the procedure discussed in Ref. 29. 

Let us consider a linear ODE of the form 
\begin{eqnarray}
\ddot{U}+\alpha\dot{U}+h_1 U = 0,\label{e9} 
\end{eqnarray}
where $\alpha $ and $\beta$ are arbitrary parameters. Now, introducing the nonlocal transformation $U = xe^{-\lambda_2}\int^{t}{ \log x(t')dt'}$, Eq. (\ref{e9}) can be transformed to the form
\begin{eqnarray}
 \ddot{X} + (\alpha-\lambda_2-2\lambda_2\ln{X})\dot{X} + h_1 X-\alpha\lambda_2X\ln{ X} +\lambda_2^2X(\ln{ X})^2 =0.\label{e10}
\label{ja22}
\end{eqnarray}
For $\alpha-\lambda_2=g_1$, (\ref{e10}) exactly matches with (\ref{e5}). Hence one can construct the general solution by following the procedure discussed in Ref 29.

\subsubsection*{Example:}
For $f=k_1$ and $\lambda_3=0$, the form of $g$ can be written from (\ref{e2}) as
\begin{eqnarray}
g=k_2-2k_1\lambda_2x,\label{ex21}
\end{eqnarray}
where $k_2=g_1+2\lambda_2\ln{k_1}$ and the form of $h(x)$ can be written from (\ref{e3}) with the help of (\ref{ex21}) as
\begin{eqnarray}
h=k_4-\lambda_2(\lambda_2+k_2)x+\lambda_2^2k_1 x^2.\label{ex22}
\end{eqnarray}
For convenience let us consider $-2k_1\lambda_2=k_3$. Then above equation can be written as
\begin{eqnarray}
h=\frac{k_3^2}{4k_1}x^2+\frac{k_3(2k_1k_2-k_3)}{4k_1^2}x+k_4.\label{ex23}
\end{eqnarray}
Now, Eq. (\ref{z1}) takes the  form
\begin{eqnarray}
\ddot{x}+k_1\dot{x}^2+(k_2+k_3x)\dot{x}+\frac{k_3^2}{4k_1}x^2+\frac{k_3(2k_1k_2-k_3)}{4k_1^2}x+k_4=0.\label{ex24}
\end{eqnarray}
The infinitesimal symmetries from (\ref{e6}) turn out to be
\begin{eqnarray}
\xi=a_1,\,\,\,\eta=\frac{c_1}{k_1}e^{\lambda_2t}.\label{ex26}
\end{eqnarray}
The infinitesimal generators read as
\begin{eqnarray}
X_1=\frac{\partial}{\partial t},\,\,\,X_2=\frac{1}{k_1}e^{\lambda_2t}\,\frac{\partial}{\partial x}.\label{ex27}
\end{eqnarray}
Introducing the transformation $x=X+A$, (\ref{ex24}) turns out to be
\begin{eqnarray}
\ddot{X}+k_1\dot{X}^2+(k_5+k_3X)\dot{X}+\frac{k_3^2}{4k_1}X^2+k_6X=0,\label{ex25}
\end{eqnarray}
 where $k_5=k_2+k_3A,\,k_6=\frac{k_3^2A}{2k_1}+\frac{k_3(2k_1k_2-k_3)}{4k_1^2}$ and $A=\frac{-2 k_1 k_2 k_3+k_3^2-\sqrt{\left(-2 k_1 k_2 k_3+k_3^2\right){}^2-16 k_1^3 k_3^2 k_4}}{2 k_1 k_3^2}$. The general solution of (\ref{ex25}) can be obtained as discussed in the previous section.


\subsection{Integrable Equation for $\bf{c=0,\,a,\,d\neq0}$}
In the subsection A of sec. VII we studied the integrable equations belonging to the case $c,\,a,d\neq0$. Now we consider the case $c=0,\,a,d\neq0$ and all its subcases. Fixing $c=0,\,a,d\neq0$ in Eq. (\ref{b1}), we have
\begin{eqnarray}
2\,\dot{d}\,G_{3x}+d\,G_{3}\,g_{x}+2\,f\,\dot{d}\,G_{3}+g\,\dot{a}-\ddot{a}=0.\label{h1}
\end{eqnarray}
Eq. (\ref{h1}) can be rewritten as
\begin{eqnarray}
g_{x}+\left(\frac{\dot{a}}{d\,G_{3}}\right)g+\frac{2\,f\,\dot{d}\,G_{3}+2\,\dot{d}\,G_{3x}-\ddot{a}}{d\,G_{3}}=0.\label{h2}
\end{eqnarray}
Since $g$ has to be a function of $x$ alone, we define 
\begin{eqnarray}
\frac{\dot{a}}{d}=\lambda_{1},\,\,\,\,\frac{\dot{d}}{d}=\lambda_{2},\qquad d\neq0.\label{h3}
\end{eqnarray}
Note that the case $d=0,\,c=0$, corresponds to the one parameter symmetry group $a=a_0$, which we have considered already. Then with the help of Eq. (\ref{h3}), Eq. (\ref{h2}) can be rewritten as
\begin{eqnarray}
g_{x}+\Bigl(\frac{\lambda_{1}}{G_{3}}\Bigl)\,g+\lambda_{2}\,\left(2\,f+\frac{2\,G_{3x}-\lambda_{1}}{G_{3}}\right)=0.\label{h4}
\end{eqnarray}
Rewriting Eq. (\ref{h4}) in terms of $f$ we get
\begin{eqnarray}
g_{x}+\lambda_{1}\,e^{\int{f\,dx}}\,(g-\lambda_{2})=0.\label{h5}
\end{eqnarray}
The above equation can be integrated to get the solution for $g$ as
\begin{eqnarray}
g=g_1 e^{-\lambda _1\int e^{\int f \, dx} \, dx}+\lambda _2,\label{h6}
\end{eqnarray}
where $g_1$ is an integration constant.

Now to explore all the equations belonging to this class we consider the possibilities $(i)\,\lambda_1\neq0,\,\lambda_2\neq0,\,(ii)\,\lambda_1\neq0,\,\lambda_2=0,\,(iii)\,\lambda_1=0,\,\lambda_2\neq0$ and $(iv)\,\lambda_1=0,\,\lambda_2=0$. We find that the cases $(ii),\,(iii)$ and $(iv)$ are either subcases of case $(i)$ or they belong to linearizable cases. Hence we need to study only the case $(i)$ separately.

\subsubsection*{$(i)\,\text{Case\,1}:\,\lambda_1\neq0,\,\lambda_2\neq0$}
Solving Eq. (\ref{h3}) for $\lambda_1\neq0$ and $\lambda_2\neq0$, we get the forms of the symmetry parameters as
\begin{eqnarray}
a=a_1+\frac{\lambda_1}{\lambda_2}d_1e^{\lambda_2t},\,\,\,d=d_1e^{\lambda_2t},\label{h10}
\end{eqnarray}
where $a_1$ and $d_1$ are constants of integration, which lead to a two parameter Lie point symmetry group. Substituting $c=0$ and using Eqs. (\ref{h3}) and (\ref{h5}) in Eq. (\ref{b2}), we get
\begin{eqnarray}
h_x+h\left(f+2\lambda_1e^{\int f \, dx}\right)+\lambda_2\left(2\lambda_2+g_1e^{-\lambda_1\int{e^{\int{fdx}}dx}}\right)=0.\label{h7}
\end{eqnarray}
Integrating (\ref{h7}) we arrive at
\begin{eqnarray}
h=e^{-\int f(x) \, dx}\left(h_1e^{-2\lambda_1\int{e^{\int{fdx}}dx}}-\frac{g_1\lambda_2}{\lambda_1}e^{-\lambda_1\int{e^{\int{fdx}}dx}}-\frac{\lambda_2^2}{\lambda_1}\right),\label{h8}
\end{eqnarray}
where $h_1$ is a constant of integration. With the forms of $g$ and $h$ [vide (\ref{h6}) and (\ref{h8})], Eq. (\ref{z1}) takes the form 
\begin{eqnarray}
\ddot{x}+f(x)\dot{x}^2+\left(g_1 e^{-\lambda _1\int e^{\int f(x) dx} dx}+\lambda _2\right)\dot{x}
+\left(h_1e^{-2\lambda_1\int{e^{\int{f(x)dx}}dx}}-\frac{g_1\lambda_2}{\lambda_1}e^{-\lambda_1\int{e^{\int{f(x)dx}}dx}}\right. \nonumber\\
\left.-\frac{\lambda_2^2}{\lambda_1}\right)e^{-\int f(x)dx}=0,\label{h9}
\end{eqnarray}
which admits the infinitesimal symmetries
\begin{eqnarray}
\xi=a_1+\frac{\lambda_1}{\lambda_2}d_1e^{\lambda_2t},\,\,\,\eta=d_1e^{\lambda_2t-\int{fdx}}.\label{h11}
\end{eqnarray}
The associated infinitesimal symmetry generators are
\begin{eqnarray}
X_1=\frac{\partial}{\partial t},\,\,\,X_2=e^{\lambda_2t}\left(\frac{\lambda_1}{\lambda_2}\,\frac{\partial}{\partial t}+e^{-\int{fdx}}\,\frac{\partial}{\partial x}\right).\label{h12}
\end{eqnarray}

Introducing now the transformation $X=\int{e^{\int{fdx}}dx}$ in (\ref{h9}) we get
\begin{eqnarray}
\ddot{X}+(\lambda_2+g_1e^{-\lambda_1X})\dot{X}-\frac{g_1\lambda_2}{\lambda_1}e^{-\lambda_1X}+h_1e^{-2\lambda_1X}
-\frac{\lambda_2^2}{\lambda_1}=0.\label{h13}
\end{eqnarray}
Interchanging the coefficients $\lambda_1$ and $\lambda_2$ for each other and then replacing the coefficient $\lambda_1$ by $\frac{\lambda_1}{\lambda_2}$, we arrive at the equation 
\begin{eqnarray}
\ddot{X}+\left(\frac{\lambda_1}{\lambda_2}+g_1e^{-\lambda_2X}\right)\dot{X}-\frac{g_1\lambda_1}{\lambda_2^2}e^{-\lambda_2X}
+h_1e^{-2\lambda_2X}-\frac{\lambda_1^2}{\lambda_2^3}=0.\label{h14}
\end{eqnarray}
The intregrability of the above equation has been discussed in Ref. 32 but for the sake of completeness we briefly discuss its here. Introducing the transformation $U = x -\frac{\lambda_1}{\lambda_2^2}t$ and $z = \frac{-\lambda_2}{\lambda_1}e^{-\frac{\lambda_1}{\lambda_2}t}$ one can rewrite (\ref{h14}) as
\begin{eqnarray}
U''+f_1e^{\lambda_2U}U'+g_1e^{-2\lambda_2U} = 0. \quad \quad ('=\frac{d}{dz})\label{ka31}
\end{eqnarray}
Eq. (\ref{ka31}) can be rewritten as
\begin{align}
U''+f_1f(U)U'+\tilde{g}_1f(U)\int f(U)dU = 0,
\label{ka32}
\end{align}
where $f(U)=e^{-\lambda_2U}$ and $\tilde{g}_1=-\lambda_2 g_1$. Eq. (\ref{ka32}) admits a time independent Hamiltonian for all values of $f_1$ and $g_1$.  The respective Hamiltonians are
\begin{eqnarray}
H=\left\{\label{ham}
\begin{array}{ll}
\frac{(r-1)}{(r-2)}p^{\frac{(r-2)}{(r-1)}}
+\frac{(r-1)f_1}{r\lambda_2}pe^{-\lambda_2U}, & \;\;f_1^2 > 4\tilde{g}_1\\
\log[p]+\frac{f_1p}{2\lambda_2}e^{-\lambda_2U},& \;\; f_1 = 4\tilde{g}_1\\
\frac{1}{2}\log\left[\frac{e^{-2\lambda_2U}}{\lambda_2^2}
\sec^2[\frac{\omega p e^{-\lambda_2U}}{-\lambda_2}]\right]+\frac{f_1}{2\lambda_2}pe^{-\lambda_2U}
,& \;\; f_1 < 4\tilde{g}_1,
\end{array}
\right. \label{E1}
\end{eqnarray}
where the canonically conjugate momentum is defined by
\begin{eqnarray}
p=\left\{
\begin{array}{ll}
{[U'+\frac{(1-r)}{r}\frac{f_1e^{-\lambda_2U}}{\lambda_2}]^{1-r}}, & \qquad f_1^2\ge 4\tilde{g_1} \\
\frac{-\lambda_2 e^{\lambda_2U}}{\omega}\tan^{-1}
[\frac{2\lambda_2U'-f_1e^{-\lambda_2U}}{2\omega e^{-\lambda_2U}}], & \qquad f_1^2 < 4\tilde{g_1}
\end{array}
\right. \label{Ab2}
\end{eqnarray}
with $r=\frac{f_1}{2g_1}(f_1\pm\sqrt{f_1^2-4\tilde{g}_1}),\;\;\omega=\frac{1}{2}\sqrt{4\tilde{g}_1-f_1^2}$. For more detail one may see Ref. 31.

\subsubsection*{Example}
For $f=-\frac{1}{x},\,g_1=k_1,\,h_1=k_2$ and $\lambda_3=0$, Eq. (\ref{h9}) takes the form
\begin{eqnarray}
\ddot{x}-\frac{1}{x}\dot{x}^2+(k_1x^{-\lambda_1}+\lambda_2)\dot{x}+k_2x^{-2\lambda_1+1}-\frac{k_1\lambda_2}{\lambda_1}x^{-\lambda_1+1}-\frac{\lambda_2^2}{\lambda_1}x=0.\label{ex51}
\end{eqnarray}
The infinitesimal symmetries for this case is
\begin{eqnarray}
\xi=a_1+\lambda_1d_1e^{\lambda_2t},\,\,\,\eta=d_1xe^{\lambda_2t},\label{ex52}
\end{eqnarray}
and the associated infinitesimal generators are
\begin{eqnarray}
X_1=\frac{\partial}{\partial t},\,\,\,X_2=e^{\lambda_2t}\left(\frac{\lambda_1}{\lambda_2}\,\frac{\partial}{\partial t}+x\,\frac{\partial}{\partial x}\right).
\end{eqnarray}
The integrability of (\ref{ex51}) can be studied in the same way as discussed in the previous section.

\section{Equivalence Transformation}
Finally we consider equivalence transformations associated with (\ref{z1}). Let us consider a set of smooth, locally one-to-one transformations $\mathcal{T}: (t,x,f,g) \longrightarrow(T,X,f_1,g_1) $ of the space $\mathbb{R}^4$ that act by the formulae
\begin{align}
T=G(t,x),\;\; X=F(t,x), \;\; f_1=H(t,x,f),\;\; g_1=L(t,x,g).
\label{ef1}
\end{align}

An equivalent transformation of Eq. (\ref{z1}) is an invertible transformation that converts Eq. (\ref{z1}) to an equation of the same form \cite{lv}
\begin{align}
\ddot{X}=-f_1(X)\dot{X}^2-g_1(X)\dot{X}-h_1(X).
\label{ef3}
\end{align}
In this case Eqs. (\ref{z1}) and (\ref{ef3}) and the functions $\{f(x)$,$g(x)\}$ and $\{f_1(X)$,$g_1(X)\}$ are equivalent.

Substituting the transformation (\ref{ef1}) into Eq. (\ref{ef3}) we get
\begin{align}
& f_1(F_t+\dot{x}F_x)^2(G_t+\dot{x}G_x)+g_1(F_t+\dot{x}F_x)(G_t+\dot{x}G_x)^2+h_1(G_t+\dot{x}G_x)^3= (F_t+\dot{x}F_x)[(G_{tt}
\nonumber\\
 & \qquad
+2\dot{x}G_{tx}+\dot{x}^2G_{xx}-G_x(f\dot{x}^2+g)]-(G_t+\dot{x}G_x)[(F_{tt}+2\dot{x}F_{tx}+\dot{x}^2F_{xx}-F_x(f\dot{x}^2+g)],
\label{ef4}
\end{align}
where the subscripts denote partial derivative with respect to that variable. Equating the coefficients of different powers of $\dot{x}^n,\;n=0,1,2,3$, we get
\begin{eqnarray}
F_xG_{xx}-G_xF_{xx}&=&f_1F_x^2G_x+g_1F_xG_x^2+h_1G_x^3 ,\label{ef5a}\\
F_tG_{xx}- G_tF_{xx}&=&Jf+f_1F_x(2F_tG_x+F_xG_t)+g_1G_x(F_tG_x+2F_xG_t)+3h_1G_tG_x^2\nonumber\\
&&+2(F_xG_{tx}-G_xF_{tx}),\label{ef5b}\\
F_xG_{tt}-G_xF_{tt}&=&Jg f_1(2F_xF_tG_t+F_t^2G_x)+g_1(F_xG_t^2+2F_tG_tG_x)+3h_1G_t^2G_x\nonumber\\
&&+2(F_tG_{tx}-G_tF_{tx}),\label{ef5c}\\
G_{tt}F_{t}-G_tF_{tt}&=&Jh+f_1F_t^2G_t+g_1F_tG_t^2+h_1G_t^3,\label{ef5d}
\end{eqnarray}
where $J=G_tF_x-F_tG_x$. Solving Eqs. (\ref{ef5a})-(\ref{ef5c}), we get the forms of the functions $f_1,g_1$ and $h_1$ as
\begin{eqnarray}
f_1&=&-\frac{1}{G_x \left(-F_x G_t+F_t G_x\right)^2}\,(2 f F_x G_t^2 G_x+F_{xx} G_t^2 G_x+4 F_x G_t G_{tx} G_x-2 f F_t G_t G_x^2\nonumber\\
&&-2 F_{tx} G_t G_x^2-g F_x G_t G_x^2-F_x G_{tt} G_x^2-2 F_t G_{tx} G_x^2+g F_t G_x^3+F_{tt} G_x^3-3 F_x G_t^2 G_{xx}\nonumber\\
&&+2 F_t G_t G_x G_{xx}),\label{ef6}\\
g_1&=&-\frac{1}{G_x^2 \left(-F_x G_t+F_t G_x\right)^2}\,(-3 f F_x^2 G_t^2 G_x-6 F_x^2 G_t G_{tx} G_x+2 f F_t F_x G_t G_x^2+2 F_{tx} F_x G_t G_x^2\nonumber\\
&&+2 g F_x^2 G_t G_x^2-2 F_t F_{xx} G_t G_x^2+2 F_x^2 G_{tt} G_x^2+2 F_t F_x G_{tx} G_x^2+f F_t^2 G_x^3+2 F_t F_{tx} G_x^3\nonumber\\
&&-2 g F_t F_x G_x^3-2 F_{tt} F_x G_x^3+3 F_x^2 G_t^2 G_{xx}-F_t^2 G_x^2 G_{xx}),\label{ef7}\\
h_1&=&-\frac{1}{G_x^2 \left(-F_x G_t+F_t G_x\right)^2}\,(f F_x^3 G_t^2 G_x+2 F_x^3 G_t G_{tx} G_x-g F_x^3 G_t G_x^2-F_x^3 G_{tt} G_x^2-f F_t^2 F_x G_x^3\nonumber\\
&&-2 F_t F_{tx} F_x G_x^3+g F_t F_x^2 G_x^3+F_{tt} F_x^2 G_x^3+F_t^2 F_{xx} G_x^3-F_x^3 G_t^2 G_{xx}).\label{ef8}
\end{eqnarray}
Substituting the values of $f_1,g_1$ and $h_1$ from Eqs. (\ref{ef6})-(\ref{ef8}) in Eq. (\ref{ef5d}) we arrive at 
\begin{eqnarray}
\left(-F_x G_t+F_t G_x\right) \left(G_t G_x \left(2 G_{tx}-g G_x\right)+G_x^2 \left(-G_{tt}+h G_x\right)+G_t^2 \left(f G_x-G_{xx}\right)\right)=0.\label{ef9}
\end{eqnarray}
As $F_x G_t-F_t G_x\neq0$, we get the condition as
\begin{eqnarray}
G_t G_x \left(2 G_{tx}-g G_x\right)+G_x^2 \left(-G_{tt}+h G_x\right)+G_t^2 \left(f G_x-G_{xx}\right)=0.\label{ef10}
\end{eqnarray}

It is difficult to solve (\ref{ef10}) to get the general form for the function $G$ and hence we have to consider the specific forms of $F$ and $G$ for simplicity. One of the possible solutions for the set of Eqs. (\ref{ef5a})-(\ref{ef5d}) is of the form
\begin{eqnarray}
F=\alpha x+\beta, \,\,\,G=\gamma t+\delta, \label{ef11}
\end{eqnarray}
where $\alpha,\beta,\gamma$ and $\delta$ are arbitrary constants. Substituting the above forms of $F$ and $G$ in Eqs. (\ref{ef5a})-(\ref{ef5d}), we get the forms for the functions $f_1$ and $g_1$ as
\begin{eqnarray}
f_1=\frac{f}{\alpha},\,\,g_{1}=\frac{ g}{\gamma},\,\,h_{1}=\frac{\alpha h}{\gamma^2}.\label{ef12}
\end{eqnarray}
Then the possible equivalence transformation is
\begin{eqnarray}
&&X=\alpha x+\beta, \,\,\,G=\gamma t+\delta,\\
&&f_1=\frac{f}{\alpha},\,\,g_{1}=\frac{ g}{\gamma},\,\,h_{1}=\frac{\alpha h}{\gamma^2}.\label{ef13}
\end{eqnarray}
With the help of (\ref{ef13}), we can write down the equivalence transformation for the examples studied in non-maximal cases. For example, the equivalence transformation for (\ref{ex14}) is given as
\begin{eqnarray}
X&=&\alpha x+\beta, \,\,\,G=\gamma t+\delta,\\
f_1&=&\frac{k_1}{X-\beta},\,\,g_1=\frac{1}{\gamma}\left(k_2+k_3\left(\frac{X-\beta}{\alpha}\right)\right),\,\nonumber\\
h_1&=&\frac{\alpha}{\gamma^2}\left(k_4\left(\frac{X-\beta}{\alpha}\right)^3+\frac{k_2k_3}{2k_1+3}\left(\frac{X-\beta}{\alpha}\right)^2+\frac{k_2^2(k_1+2)}{(2k_1+3)^2}\,\frac{X-\beta}{\alpha}\right),
\end{eqnarray}
whereas for (\ref{ex24}) the transformation is
\begin{eqnarray}
X&=&\alpha x+\beta, \,\,\,G=\gamma t+\delta,\\
f_1&=&\frac{k_1}{\alpha},\,\,g_1=\frac{1}{\gamma}\left(k_2+k_3\left(\frac{X-\beta}{\alpha}\right)\right),\,\nonumber\\
h_1&=&\frac{\alpha}{\gamma^2}\left(\frac{k_3^2}{4k_1}\left(\frac{X-\beta}{\alpha}\right)^2+\frac{k_3(2k_1k_2-k_3)}{4k_1^2}\,\left(\frac{X-\beta}{\alpha}\right)+k_4\right).
\end{eqnarray}
The transformation for (\ref{ex51}) can be written as
\begin{eqnarray}
X&=&\alpha x+\beta, \,\,\,G=\gamma t+\delta,\\
f_1&=&-\frac{1}{X-\beta},\,\,g_1=\frac{1}{\gamma}\left(k_1\left(\frac{X-\beta}{\alpha}\right)^{-\lambda_1}+\lambda_2\right),\nonumber\\
h_1&=&\frac{\alpha }{\gamma^2}\left(k_2\left(\frac{X-\beta}{\alpha}\right)^{1-2\lambda_1}-\frac{k_1\lambda_2}{\lambda_1}\left(\frac{X-\beta}{\alpha}\right)^{1-\lambda_1}-\frac{\lambda_2^2}{\lambda_1}\left(\frac{X-\beta}{\alpha}\right)\right).
\end{eqnarray}

\section{Conclusion}
\label{sec9}
In this paper we have succeeded to deduce all the Lie point symmetry groups associated with the mixed quadratic-linear Li$\acute{e}$nard type Eq. (\ref{z1}) by developing a self consistent procedure. We have systematically identified and classified all those equations which admit one, two and eight parameter symmetry groups. We have identified the general form of the ODE which can be linearized under Lie point transformations. The explicit form of the equation as well as the associated eight Lie point symmetries which it admits have been presented. We have also reported the linearizing transformation, the first integrals and general solution of this nonlinear ODE. We have observed that Eq. (\ref{z1}) does not admit a three parameter symmetry for any nonzero forms of $f,g$ and $h$. As far as two parameter symmetry group is concerned we have brought out three different forms of nonlinear ODEs. We have given the explicit time independent Hamiltonians associated with these three equations and proved their Liouville integrability.  

One can also transform the general linearizable nonlinear ODE (\ref{z2.1}) to another nonlinear ODE which is of the form (Li$\acute{e}$nard type equation), $\ddot{X}+(k_1X+k_2)\dot{X}+\frac{k_1^2}{9}X^3+\frac{k_1k_2}{3}X^2+k_3X+k_4=0$ through the transformation $X=\int{e^{\int{f(x)dx}}dx}$. This transformed nonlinear ODE is yet another linearizable equation. Similarly in the category of integrable equations also while transforming them into some known equations we have noticed that they fall again into the Li$\acute{e}$nard type equation (vide (\ref{d6.1}), (\ref{e8}) and (\ref{h14})).  Thus we conclude that both the linearizable and integrable cases that admit Lie point symmetries in the family of Eq. (\ref{z1}) are all intimately related to the Li$\acute{e}$nard equation (\ref{intr1}). There may be other integrable/linearizable equations which may not go into Li$\acute{e}$nard equation as in the case of equations admitting one parameter Lie symmetries, see for example Mathews-Lakshmanan oscillator equation which admits $\lambda-$symmetries. It is clear that Lie point symmetries, except for translation symmetry, are absent in those equations and some kind of generalized symmetries should exist in these cases. One has to look for such generalized symmetries (which are mentioned in the introduction) in order to establish the integrability of the underlying equation. It will be a challenging problem to explore such symmetries systematically.   

\section{Acknowledgments}
AKT and SNP are grateful to the Centre for Nonlinear Dynamics, Bharathidasan University, Tiruchirappalli, for warm hospitality. The work of SNP forms part of a Department of Science and Technology, Government of India, sponsored research project. The work of MS forms part of a research project sponsored by UGC. The work forms part of a Department of Science and Technology, Govt. of India IRHPA project and a DST Ramanna Fellowship project of ML. He also acknowledges the financial support provided through a DAE Raja Ramanna Fellowship.

\appendix
\begin{appendix}
\section{Relationship among the functions $f,\,g$ and $h$}
Integrating Eqs. (\ref{a8})-(\ref{a11}) consistently one can obtain the infinitesimals $\xi$ and $\eta$ in terms of the arbitrary functions
 $f,\,g$ and $h$. Eventhough  Eq. (\ref{a8}) can be integrated to yield $\xi$, with two unknown arbitrary functions, the explicit form of these 
two unknown arbitrary functions as well as the other infinitesimal $\eta$ are difficult to obtain explicitly from the rest of the determining equations. 
To determine the infinitesimals $\xi$ and $\eta$ uniquely from the determining Eqs. (\ref{a8})-(\ref{a11}), we formulate two integrability 
conditions by imposing the compatibility between these four equations. By considering each one of the integrability 
conditions separately, we then integrate the determining Eqs. (\ref{a8})-(\ref{a11})  and explore the Lie point symmetries of (\ref{z1}).

\subsection*{$(i)$\,First auxiliary equation}
To start with we seek the compatibility between Eqs. (\ref{a9}) and (\ref{a10}). We then simplify this equation appropriately so that the resultant equation (which we call as first auxiliary equation) involves the infinitesimals $\xi$ and $\eta$ and their first derivative only. 

Differentiating Eq. (\ref{a9}) with respect to $t$,
\begin{eqnarray}
\eta_{xxt}+f \eta_{xt}+f_{x}\eta_{t}-2 \xi_{txt}+2 g \xi_{xt}=0.\label{a12}
\end{eqnarray}
Next, then replacing the term $\eta_{xt}$ in the resultant equation by substituting (\ref{a10}) in it, one gets
\begin{eqnarray}
2\eta_{xxt}-f \eta\,  g_{x}-3f h\,\xi_{x}+2\,\left(f_x-f^2\right)\eta_{t}-f g\,\xi_{t}+4g\,\xi_{xt}+f\xi_{tt}-4\xi_{txt}=0.\label{a13}
\end{eqnarray}
Differentiating Eq. (\ref{a10}) with respect to $x$ once, we get  
\begin{eqnarray}
2\eta_{txx}+\eta  g_{xx}+g_x \eta _x+2 f_x \eta _t+2 f \eta _{tx}+g_x \xi _t+g \xi _{tx}-\xi _{ttx}+3 h_x \xi _x+3 h \xi _{xx}=0.\label{a14}
\end{eqnarray}
Substituting Eqs. (\ref{a8}) and (\ref{a10}) in Eq. (\ref{a14}) for $\xi_{xx}$ and $\eta_{xt}$ respectively and simplifying the resultant equation, we obtain the relation
\begin{eqnarray}
2\eta_{txx}+\eta \left(g_{xx}-fg_x\right)+g_x\eta_x+3\,h_x\,\xi_x+2\left(f_x-f^2\right)\eta_t+\left(g_x-fg\right)\xi_t+g\,\xi_{tx}\nonumber\\
+f\xi_{tt}-\xi_{ttx}=0.\label{a15}
\end{eqnarray}
Eliminating $\eta_{xxt}$ from (\ref{a13}) and (\ref{a15}), we arrive at
\begin{eqnarray}
\eta  g_{xx}+g_x \eta _x+3\left(h_x+f h\right) \xi _x+g_x \xi _t-3 g \xi _{tx}+3 \xi _{ttx}=0.\label{a16}
\end{eqnarray}

The expression which we obtained from the compatibility between Eqs. (\ref{a9}) and (\ref{a10}) contains the terms $\xi_{tx}$ and $\xi_{ttx}$. Now, we express these two terms in terms of first derivatives. To achieve this we differentiate (\ref{a12}) with respect to $x$ and obtain
\begin{eqnarray}
\eta_{xxtx}+f_{xx}\eta_t+f_x\eta_{xt}+f \eta_{xtx}+f_x\eta_{tx}-2 \xi_{txtx}+2 g_x \xi_{xt}+2 g \xi_{xtx}=0.\label{a17}
\end{eqnarray}
We again replace the terms $\xi_{xxt}$ and $\xi_{txtx}$ which appear in (\ref{a17}), in terms of their lower order derivatives by using Eq. (\ref{a8}), and then $\eta_{xt}$ by (\ref{a10}) and finally $\eta_{xtx}$ by (\ref{a13}) so that Eq. (\ref{a17}) now simplifies to 
\begin{eqnarray}
2\eta_{xxtx}-\eta\,g_x \left(2f_x-f^2\right)-3\,h\,\left(2 f_x-f^2\right) \xi _x+2\left(f_{xx}-3 f f_x+f^3\right) \eta_t\nonumber\\
-g\left(2f_x-f^2\right) \xi_t+4 g_x \xi_{xt}+\left(2f_x-f^2\right) \xi_{tt}=0.\label{a18}
\end{eqnarray}

Differentiating (\ref{a14}) with respect to $x$, we get
\begin{eqnarray}
2\eta_{txxx}+\eta  g_{xxx}+2 g_{xx}\eta _x+3 h_{xx}\xi _x+g_x\eta _{xx}+6 h_x\xi _{xx}+3 h\xi _{xxx}+2 f_{xx}\eta _t+g_{xx}\xi _t\nonumber\\
+4 f_x\eta _{tx}+2 g_x\xi _{tx}+2 f\eta _{txx}+g\xi _{txx}-\xi_{ttxx}=0.\label{a19}
\end{eqnarray}
Substituting Eqs. (\ref{a8})-(\ref{a11}) and (\ref{a15}) in the above Eq. (\ref{a19}) and simplifying the later, we find
\begin{eqnarray}
2\eta_{txxx}+\eta  \left(g_{xxx}-f g_{xx}-3f_x g_x+f^2 g_x\right)+\left(3h_{xx}-3 h f_x+3f h_x-2g g_x+3f^2 h\right) \xi_x\nonumber\\
+2\left(g_{xx}-f g_x\right) \eta_x+2\left(f_{xx}-3 f f_x+f^3\right) \eta_t+\left(g_{xx}-2g f_x-f g_x+f^2 g\right) \xi_t+4 g_x \xi_{tx}\nonumber\\
+\left(2f_x-f^2\right) \xi _{tt}=0.\label{a20}
\end{eqnarray}
Eliminating the variable $\eta_{xxtx}$ in (\ref{a18}) with the help of (\ref{a20}), we arrive at
\begin{eqnarray}
 \eta  \left(g_{xxx}-f g_{xx}-f_x g_x\right)+2\left(g_{xx}-f g_x\right) \eta _x+\left(3 h_{xx}+3 h f_x+3 f h_x-2 g g_x\right) \xi _x\nonumber\\
+\left(g_{xx}-f g_x\right) \xi _t=0.\label{a21}
\end{eqnarray}

Note that Eq. (\ref{a21}) which comes out from the compatibility of the Eqs. (\ref{a9}) and (\ref{a10}), with the help of appropriate usage other two Eqs. (\ref{a8}) and (\ref{a11}) in them, involves the infinitesimals $\xi,\,\eta$ and their first derivatives $\eta_x,\,\xi_x$ and $\xi_t$ alone. The coefficients of the derivatives of $\xi$ and $\eta$ now turn out to be functions of $f,\,g,\,h$ and their derivatives. 

To proceed further, we rewrite Eq. (\ref{a21}) in the form
\begin{eqnarray}
PL_1+ \xi_xL_2+\eta L_{1x} =0,\label{a33}
\end{eqnarray}
where 
\begin{eqnarray}
&&L_1=g_{xx}-f g_x,\label{a31}\\
&&L_2=3 h_{xx}+3 h f_x+3 f h_x-2 g g_x,\label{a32}
\end{eqnarray}
with $P=2\eta_x+\xi_t$. We call Eq. (\ref{a32}) as the first auxiliary equation.

\subsection*{$(ii)$\,Second auxiliary equation}
We construct the second auxiliary equation by considering the compatibility between Eqs. (\ref{a9}) and (\ref{a11}). We then simplify this equation augmenting another equation which comes out from the compatibility of Eqs. (\ref{a10}) and (\ref{a11}) so that the resultant expression involves only $\xi$ and $\eta$ and does not contain any second derivatives. 

Now, let us differentiate Eq. (\ref{a11}) once with respect to $x$, which in turn yields
\begin{eqnarray}
\eta _{ttx}+\eta  h_{xx}-h \eta _{xx}+g_x \eta _t+2 h_x \xi _t+g \eta _{tx}+2 h \xi _{tx}=0.\label{a22}
\end{eqnarray}
Repeating the differentiation one more time, we find
\begin{eqnarray}
\eta_{ttxx}+\eta  h_{xxx}+h_{xx} \eta _x-h_x \eta _{xx}-h \eta _{xxx}+g_{xx} \eta _t+2 h_{xx} \xi _t+2 g_x \eta _{tx}+4 h_x \xi _{tx}+g \eta _{txx}\nonumber\\
+2 h \xi _{txx}=0.\label{a23}
\end{eqnarray}
Substituting Eqs. (\ref{a8})-(\ref{a11}) in (\ref{a23}) and then replacing $\eta_{xxt}$ by (\ref{a13}) and simplifying the resultant equation, we obtain
\begin{eqnarray}
2\eta_{ttxx}+\eta  \left(2h_{xxx}+2h f_{xx}+2f_x h_x-2f h f_x+ f g g_x-2g_x^2\right)+\left(4 h_{xx}-2g g_x+f g^2\right) \xi _t\nonumber\\
+\left(4 g h_x-2h g_x+3 f g h\right) \xi _x+2\left(g_{xx}-g f_x-2 f g_x+f^2 g\right) \eta _t+4\left(h_x + f h-g^2\right) \xi _{tx}\nonumber\\
+2\left(h_{xx}+2 h f_x+f h_x-f^2 h\right) \eta _x+\left(2g_x -f g\right) \xi _{tt}+4 g \xi _{txt}=0.\label{a24}
\end{eqnarray}

Next, we consider Eq. (\ref{a12}) and differentiate it with respect to $t$, so that one has
\begin{eqnarray}
\eta_{xxtt}+f_x \eta_{tt}+f \eta_{xtt}+2 g \xi_{xtt}-2 \xi_{txtt}=0.\label{a25}
\end{eqnarray}
Now let us replace the term $\eta_{tt}$ in the above Eq. (\ref{a25}) by (\ref{a11}) and $\eta_{xtt}$ by (\ref{a22}). As a consequence Eq. (\ref{a25}) can now be brought to the form 
\begin{eqnarray}
2\eta_{xxtt}-\eta \left(2f h_{xx}+2f_x h_x+2 h f f_x- f g g_x\right)- f \,g\, h\, \xi _x-2\left(g f_x+f g_x-f^2 g\right)\eta _t\nonumber\\
 +2\,h\,\left(f_x-f^2\right) \eta _x-\left(4 h f_x+4 f h_x-f g^2\right) \xi _t- f g \xi _{tt}+4 g \xi _{xtt}-4 \xi _{txtt}=0.\label{a26}
\end{eqnarray}
The fourth derivative of $\eta$, namely, $\eta_{ttxx}$, can be eliminated from (\ref{a24}) with the help of (\ref{a26}). The result shows that
\begin{flalign}
\eta  \left(h_{xxx}+h f_{xx}+f h_{xx}+2 f_x h_x-{g_x}^2\right)+\left(2 g h_x+2 f g h-h g_x\right) \xi _x+\left(g_{xx}-f g_x\right) \eta _t\nonumber\\
+\left(h_{xx}+h f_x+f h_x\right) \eta _x+\left(2 h_{xx}+2 h f_x+2 f h_x-g g_x\right) \xi _t+2\left(h_x+f h-g^2\right) \xi _{tx}\nonumber\\
+g_x \xi _{tt}+2 \xi _{txtt}=0.\label{a27}
\end{flalign}

The equation which arises from the compatibility of (\ref{a9}) and (\ref{a11}) contains both second and higher derivative terms. As in the previous case here also we try to formulate an auxiliary equation which involves first derivatives only. To achieve this task, we express the terms $\xi_{tt},\,\xi_{tx}$ and $\xi_{txtt}$ which appear in (\ref{a27}) in terms of their lower order derivatives. We differentiate Eq. (\ref{a14}) with respect to $t$ and obtain an expression
\begin{eqnarray}
2\eta_{txxt}+ g_{xx} \eta_t+g_x \eta_{xt}+3 h_x \xi_{xt}+3 h \xi_{xxt}+2 f_x \eta_{tt}+g_x \xi_{tt}+2 f \eta_{txt}+g \xi_{txt}-\xi_{ttxt}=0.\label{a28}
\end{eqnarray}
Substituting Eqs (\ref{a8})-(\ref{a10}) and (\ref{a22}) in (\ref{a28}) and simplifying the later equation, we get
\begin{eqnarray}
4 \eta_{txxt}-\eta  \left(4f h_{xx}+4f_x\, h_x+4f\, h\, f_{x}-2 f\, g \,g_{x}+ {g_{x}}^2\right)+4\left(h f_x-f^2 h\right) \eta_x+2 g \xi_{txt}\nonumber\\
+2\left(g_{xx}-3 f g_x-2g f_x+2f^2 g\right) \eta_t-\left(8 f h_x+8 h f_x+ g g_x-2 f g^2\right) \xi _t-2 \xi_{ttxt}\nonumber\\
-h\,\left(3 g_x+2 f g\right) \xi_x+6\left(h_x+f h\right) \xi_{tx}+\left(3 g_x-2fg\right) \xi_{tt}=0.\label{a29}
\end{eqnarray}
To eliminate the term $\eta_{ttxx}$ in (\ref{a29}) we substitute (\ref{a24}) in the former. The result yields
\begin{eqnarray}
\eta  \left(4 h_{xxx}+4 f h_{xx}+4 h f_{xx}+8 f_x h_x-3 {g_x}^2\right)+\left(8 h_{xx}+8h f_x+8f h_x-3 g g_x\right) \xi _t\nonumber\\
+\left(8 g h_x+8 f g h-h g_x\right) \xi _x+2\left( g_{xx}-f g_x\right) \eta _t+4\left(h_{xx}+h f_x+f h_x\right) \eta _x\nonumber\\
+2\left(h_x+f h-4 g^2\right) \xi _{tx}+g_x \xi _{tt}+6 g \xi _{txt}+2 \xi _{ttxt}=0.\label{a30a}
\end{eqnarray}

Now, subtracting (\ref{a27}) from (\ref{a30a}) we can remove the terms $\xi_{tt}$ and $\xi_{txtt}$ but the resultant equation involves the terms $\xi_{txt}$. To eliminate this term we multiply (\ref{a16}) by $2g$ and add the later one to the subtracted equation which in turn cancels the term $\xi_{txt}$. The final expression obtained in this process reads 
\begin{eqnarray}
\hspace{-.2in}\eta  \left(3 h_{xxx}+3 f h_{xx}+3 h f_{xx}+6 f_x h_x-2 {g_x}^2-2 g g_{xx}\right)+\left(3 h_{xx}+3 h f_x+3 f h_x-2 g g_x\right) \eta _x\nonumber\\
+\left(g_{xx}-f g_x\right) \eta _t+2\left(3 h_{xx}+3 h f_x+3 f h_x-2 g g_x\right) \xi _t=0.\label{a30}
\end{eqnarray}

Eq. (\ref{a30a}) can be rewritten in a compact form 
\begin{eqnarray}
 \eta _tL_ 1 +  QL_2 + \eta  L_ {2x} =  0,\label {a49}
\end{eqnarray}
where $L_1$ and $L_2$ are given in (\ref{a31}) and (\ref{a32}) and $Q=\eta _x + 2\xi _t$. We call Eq. (\ref{a49}) as the second auxiliary equation. This auxiliary equation which arises from the compatibility of both (\ref{a9}) and (\ref{a11}) as well as  (\ref{a10}) and (\ref{a11}) involves only the infinitesimals $\xi$ and $\eta$ and their first derivatives. Interestingly, we observe that the compatibility conditions have now been written in a neater form and more importantly with two common coefficients, namely $L_1$ and $L_2$. Now, we can solve these two auxiliary equations and formulate the integrability conditions for the determining equations.

\subsection*{(iii)\,Integrability conditions}
We consider the auxiliary equations as two algebraic equations for the unknowns $L_1$ and $L_2$. In other words we try to remove the terms $L_{1x}$ and $L_{2x}$  which appear in these two equations. We then solve the resultant algebraic equations and establish the integrability condition for the existence of symmetries. 

Differentiating (\ref{a33}) with respect to $t$, we get
\begin{eqnarray}
P_{t}L_1+\xi_{tx} L_2+\eta_{t} L_{1x}=0,\label{a34}
\end{eqnarray}
and replacing the term $L_{1x}$ by using (\ref{a33})again, Eq. (\ref{a34}) can be brought to the form
\begin{eqnarray}
\left(\frac{P}{\eta}\right)_tL_1+\left(\frac{\xi_x}{\eta}\right)_tL_2=0.\label{a36}
\end{eqnarray}
Similarly, differentiating Eq (\ref{a34}) with respect to $t$,
\begin{eqnarray}
P_{tt}L_1+\xi_{ttx} L_2+\eta_{tt} L_{1x}=0,\label{a37}
\end{eqnarray}
and then replacing $L_{1x}$ in the resultant equation by (\ref{a33}), we obtain another equation which involves only $L_1$ and $L_2$ that is
\begin{eqnarray}
\left(P_{tt} \eta-P\eta_{tt}\right)L_1+\left(\xi_{ttx} \eta-\xi_x \eta_{tt}\right) L_2=0.\label{a38} 
\end{eqnarray}
Let us also rewrite Eq. (\ref{a38}) to be in the same form as that of (\ref{a36}). Recalling the derivatives 
\begin{eqnarray}
&&\left(\frac{P}{\eta}\right)_t=\frac{P_t \eta-P \eta_t}{\eta^2}.\label{a39}\\
\text{and}
&&\left(\frac{P}{\eta}\right)_{tt}=\frac{(P_{tt} \eta-P \eta_{tt})}{\eta^2}-\frac{2\eta_t(P_t\eta-P\eta_t)}{\eta^3},\label{a40}
\end{eqnarray}
we can re-express the first term in (\ref{a38}) as
\begin{eqnarray}
P_{tt}\eta-P\eta_{tt}=\eta^2\left(\left(\frac{P}{\eta}\right)_{tt}+\frac{2\eta_t}{\eta}\left(\frac{P}{\eta}\right)_{t}\right).\label{a41}
\end{eqnarray}
Similarly, we can also express the second term in (\ref{a38}) in the form
\begin{eqnarray}
\xi_{ttx}\eta-\xi_x\eta_{tt}=\eta^2\left(\left(\frac{\xi_x}{\eta}\right)_{tt}+\frac{2\eta_t}{\eta}\left(\frac{\xi_x}{\eta}\right)_t\right).\label{a42}
\end{eqnarray}
With the help of (\ref{a41}) and (\ref{a43}), Eq. (\ref{a38}) can be rewritten as
\begin{eqnarray}
\left(\left(\frac{P}{\eta}\right)_{tt}+\frac{2\eta_t}{\eta}\left(\frac{P}{\eta}\right)_t\right)L_1+\left(\left(\frac{\xi_x}{\eta}\right)_{tt}+\frac{2\eta_t}{\eta}\left(\frac{\xi_x}{\eta}\right)_t\right)L_2=0.\label{a43}
\end{eqnarray}
Upon solving Eqs. (\ref{a36}) and (\ref{a43}), we find that the solution exists under two conditions, namely,  
\begin{eqnarray}
&&(i)\,L_1,\,L_2\neq0,\,\,\,M=0,\label{a44a}\\
&&(ii)\,L_1=0,\,\,\,L_2=0,\,\,\,M\neq0,\label{a44b}
\end{eqnarray}
where 
\begin{eqnarray}
M=\left(\frac{P}{\eta}\right)_{tt}\left(\frac{\xi_x}{\eta}\right)_t-\left(\frac{\xi_x}{\eta}\right)_{tt}\left(\frac{P}{\eta}\right)_t.\label{a44.1}
\end{eqnarray}

By recalling the second auxiliary equation (vide Eq. (\ref{a49})) and proceeding in the same manner as we did above for Eq. (\ref{a33}), we can derive the following two equations from (\ref{a49}), namely
\begin{eqnarray}
&&\left(\frac{\eta_t}{\eta}\right)L_1+\left(\frac{Q}{\eta}\right)L_2=0,\label{1a1}\\
&&\left(\left(\frac{\eta_t}{\eta}\right)_{tt}+\frac{2\eta_t}{\eta}\left(\frac{\eta_t}{\eta}\right)_t\right)L_1
+\left(\left(\frac{Q}{\eta}\right)_{tt}+\frac{2\eta_t}{\eta}\left(\frac{Q}{\eta}\right)_t\right)L_2=0.\label{1a2}
\end{eqnarray}
Upon solving these two equations, (\ref{1a1}) and (\ref{1a2}), we find that the following two conditions should be fulfilled in order to have non-trivial solutions, namely
\begin{eqnarray}
&&(i)\,L_1,\,L_2\neq0,\,\,\,N=0\label{a51a}\\
\text{and} 
&&(ii)\,L_1=0,\,\,\,L_2=0,\,\,\,N\neq0,\label{a51b}
\end{eqnarray}
where $N$ is given by
\begin{eqnarray}
N=\left(\frac{\eta_t}{\eta}\right)_{tt}\left(\frac{Q}{\eta}\right)_t-\left(\frac{Q}{\eta}\right)_{tt}\left(\frac{\eta_t}{\eta}\right)_t.\label{a51.1}
\end{eqnarray}

Analyzing Eqs. (\ref{a44a}), (\ref{a44b}), (\ref{a51a}) and (\ref{a51b}), we conclude that one can integrate the determining Eqs. (\ref{a8})-(\ref{a11}) and obtain for the ODE (\ref{z1}) nontrivial symmetries under the following two conditions, namely $(i)\,L_1=0,\,L_2=0,\,M\neq0 ,\, N\neq0$ and $(ii)\,L_1\neq0,\,L_2\neq0,\,M=0 ,\,N=0$. We call these two conditions as the integrability conditions for the system of determining equations. The first condition says that both the expressions which involve arbitrary functions $f,g$ and $h$ (say $L_1$ and $L_2$) should vanish whereas none of the expressions which interconnects the infinitesimals $\xi$ and $\eta $ should vanish. From the second condition we infer that this should be in the opposite way. The expressions ($M$ and $N$) which involves the infinitesimals should vanish whereas the expression which involves the arbitrary functions should not vanish. We note here that the conditions are independent in the sense one can integrate the determining equations in both the conditions to obtain nontrivial solutions.   

In the following, we analyze these two conditions in some detail.

\section{The Case $L_1=0,\,L_2=0$ and $M,\,N\neq0$.}
Now, we analyze first condition, that is $L_1=0,\,L_2=0$ and $M,\,N\neq0$. It has been proved that the linearization of a scalar second order nonlinear ODE, $\ddot{x}+f(t,x,\dot{x})=0$, by point transformation will result in a cubic polynomial in first derivative \cite{ibra}, 
\begin{equation}
\ddot{x}=A(t,x)\dot{x}^{3}+B(t,x)\dot{x}^{2}+C(t,x)\dot{x}+D(t,x),\label{appe1}
\end{equation}
where the coefficients $A,\,B,\,C$ and $D$ should satisfy the following two conditions, namely,
\begin{eqnarray}
3A_{tt}+3A_{t}C-3A_{x}D+3AC_{t}+C_{xx}-6AD_{x}+BC_{x}-2BB_{t}-2B_{tx}=0,\label{appe2.1}\\
6A_{t}D-3B_{x}D+3AD_{t}+B_{tt}-2C_{tx}-3BD_{x}+3D_{xx}+2CC_{x}-CB_{t}=0,\label{appe2}
\end{eqnarray}
where the suffix refers to partial derivatives. By comparing our case (\ref{z1}) with (\ref{appe1}), we find that the linearizability criteria (\ref{appe2.1}) and (\ref{appe2}) reduces to 
\begin{eqnarray}
&&g_{xx}-fg_{x}=0,\label{appe3}\\
&&3h_{xx}+3 h f_x+3 f h_x-2 g g_x=0,\label{appe4}
\end{eqnarray}
\textit{which is exactly equivalent to $L_1=0$ and $L_2=0$}. Thus we observe that the first integrability condition which we have derived in the previous section is nothing but the linearizability condition of the undertaken nonlinear ODE (\ref{z1}) under point transformation. It has also been proved  that the linearizable second order ODE admits maximal number of Lie point symmetries. \textit{Thus we conclude that whenever the given functions $f,\,g$ and $h$ satisfy the relations (\ref{appe3}) and (\ref{appe4}) the symmetry determining equations (\ref{a8})-(\ref{a11}) can be integrated to give maximal number of symmetries}. This is also confirmed by the non-vanishing of $M$ and $N$.

\section{The Case $L_1\neq0,\, L_2\neq0\, \text{and} \,M=0,\,N=0$. }
Now, we analyze the second condition, that is, $L_1\neq0,L_2\neq0$ and $M=0,N=0$. Since we have $L_1\neq0$ and $L_2\neq0$ one may preassume that the determining equations can be integrated to yield lesser parameter Lie point symmetries only. However, here we are interested in understanding the implications  of the conditions $M=0$ and $N=0$. For this purpose we investigate these two conditions in detail. 

To begin with let us consider the condition $M=0$ which is equivalent to 
\begin{eqnarray}
\left(\frac{P}{\eta}\right)_{tt}\left(\frac{\xi_x}{\eta}\right)_t=\left(\frac{\xi_x}{\eta}\right)_{tt}\left(\frac{P}{\eta}\right)_t.\label{a46}
\end{eqnarray}
Eq. (\ref{a46}) can also be written as
\begin{eqnarray}
\frac{\partial}{\partial t}\left(\log\left[\left(\frac{P}{\eta}\right)_t\right]\right)=\frac{\partial}{\partial t}\left(\log\left[\left(\frac{\xi_x}{\eta}\right)_t\right]\right),\label{a46.1}
\end{eqnarray}
where $P=2\eta_x+\xi_t$. Integrating (\ref{a46.1}) and substituting the form of $P$ in it, we get
\begin{eqnarray}
2\eta_x+\xi_t=c_1(x)\xi_x+c_2(x)\eta,\label{a47}
\end{eqnarray}
where $c_1$ and $c_2$ are arbitrary functions of $x$. 

Repeating the calculation for $N=0$, in the same manner, we end up at
\begin{eqnarray}
\eta_t=c_3(x)(\eta_x+2\xi_t)+c_4(x)\eta,\label{a53}
\end{eqnarray}
where $c_3$ and $c_4$ are functions of $x$. With the help of (\ref{a47}) the above Eq. (\ref{a53}) can be rewritten as 
\begin{eqnarray}
\eta_t=\eta \left(\frac{1}{2} c_2 c_3+c_4\right)+\frac{1}{2} c_1 c_3 \xi_x +\frac{3}{2} c_3 \xi_t.\label{ap1}
\end{eqnarray}
Now, we seek a compatibility between Eqs. (\ref{a47}) and (\ref{ap1}) so that the resultant equation admits only one variable, say $\xi$ (which can be determined from (\ref{sym1})). Doing so, we find
\begin{eqnarray}
\left(c_2 c_{3x}+c_3 c_{2x}+2 c_{4x}\right)\eta+\left(c_1c_4+c_1 c_{3x}+ c_3c_{1x}\right) \xi_x+\left(3c_{3x} -2c_2 c_3 -c_4\right)\xi_t+ c_1 c_3 \xi_{xx}\nonumber\\
 +\left(3 c_3- c_1\right) \xi_{tx}+ \xi_{tt}=0.\label{ap4}
\end{eqnarray}

Differentiating Eq. (\ref{ap4}) with respect to $x$, we obtain an equation of the form
\begin{eqnarray}
A_{1}\eta+A_2\xi_x+A_3\xi_{xx}+A_4\xi_{xxx}+A_5\xi_t+A_6\xi_{tx}+A_7\xi_{txx}+2\xi_{ttx}=0,\label{ap5}
\end{eqnarray}
where the coefficients ${A_i}^{'}s,\,i=1,2,....,7,$ are functions of $x$ alone and are found to be
\begin{eqnarray}
A_1&=&2c_2c_{3xx}+2c_3 c_{2xx}+4 c_{2x} c_{3x}+4c_{4xx}+ c_2c_3 c_{2x}+c_2^2c_{3x}+2c_2c_{4x},\nonumber\\
A_2&=&4c_1c_{4x}+2c_4c_{1x}+2c_1c_{3xx}+2c_3c_{1xx}+c_1c_3c_{2x}+ c_1c_2c_{3x}+4c_{1x}c_{3x},\nonumber\\
A_3&=&2 c_1c_4+4 c_1c_{3x}+4 c_3 c_{1x},\,\,\,A_4=2c_1c_3,\nonumber\\
A_5&=&-5 c_3c_{2x}-5 c_2 c_{3x}-4 c_{4x}+6 c_{3xx},\nonumber\\
A_6&=&-4 c_2 c_3-2 c_4-2 c_{1x}+12 c_{3x},\,\,\,A_7=6c_3-2c_1.\label{ap6}
\end{eqnarray}

We recall here that the infinitesimal $\xi$ has two arbitrary functions in it, $a(t)$ and $b(t)$. Further we do not know the expression of $\eta$ as well. So we conclude that Eq. (\ref{ap5}) contains three unknowns, namely $\eta,\,a$ and $b$. We aim to retain only one unknown, namely $b$ alone, in this equation after some differentiation/modification. 

To achieve this task we differentiate (\ref{ap5}) one more time with respect to $x$ and obtain
\begin{eqnarray}
\hspace{-.2in}\eta \left(A_1 c_2+2 A_{1x}\right)+\left(A_1 c_1+2 A_{2x}\right) \xi_x+\left(2 A_2+2 A_{3x}\right) \xi_{xx}+\left(2 A_3+2 A_{4x}\right) \xi_{xxx}+2A_4\xi_{xxxx}\nonumber\\
+\left(-A_1+2A_{5x}\right)\xi_t+\left(2A_5+2 A_{6x}\right) \xi_{tx}+\left(2 A_6+2 A_{7x}\right) \xi_{txx}+2 A_7 \xi_{txxx}+4\xi_{ttxx}=0.\label{ap7}
\end{eqnarray}
Since the term $\xi_t$ contains two unknowns, $a$ and $b$, we try to eliminate this term. We eliminate $\xi_t$ from (\ref{ap7}) by using (\ref{ap5}). The resultant equation reads
\begin{eqnarray}
\eta B_1+B_2 \xi_x+B_3 \xi_{xx}+B_4 \xi_{xxx}+B_5 \xi_{xxxx}+B_6 \xi_{tx}+B_7\xi_{txx}+B_8 \xi_{txxx}+B_9 \xi_{ttx}\nonumber\\
+B_{10}\xi_{ttxx}=0,\label{ap8}
\end{eqnarray}
where the functions ${B_i}^{'}s,\,i=1,2,....,10,$ are functions of $x$ alone and are found to be
\begin{eqnarray}
&&B_1=A_1(A_1-2A_{5x})+A_5 (A_1c_2+2A_{1x}),\,\,B_2=A_2(A_1-2A_{5x})+A_5 (A_1c_1+2A_{2x}),\nonumber\\
&&B_3=A_3(A_1-2A_{5x})+2A_5 (A_2+A_{3x}),\,\,B_4=A_4(A_1-2A_{5x})+2A_5(A_3+A_{4x}),\nonumber\\
&&B_5=2A_4A_5,\,\,B_6=A_6(A_1-2A_{5x})+2A_5 (A_5+A_{6x}),\nonumber\\
&&B_7=A_7(A_1-2A_{5x})+2A_5 (A_6+A_{7x}),\,\,B_8=2A_5A_7,\nonumber\\
&&B_9=2 A_1-4 A_{5x},\,\,\,B_{10}=4A_5.\label{ap9}
\end{eqnarray}

One may note that Eq. (\ref{ap8}) is free from the arbitrary function $a$. To eliminate the other unknown $\eta$ in (\ref{ap8}), we differentiate it with respect to $x$ and get
\begin{eqnarray}
C_1\eta +C_2\xi_x+C_3\xi_{xx}+C_4\xi_{xxx}+C_5\xi_{xxxx}+C_6\xi_{xxxxx}+C_7\xi_{tx}+C_8\xi_{txx}+C_9\xi_{txxx}\nonumber\\
+C_{10}\xi_{txxxx}+C_{11}\xi_{ttx}+C_{12}\xi_{ttxx}+C_{13}\xi_{ttxxx}=0,\label{ap10}
\end{eqnarray}
where ${C_i}^{'}s,\,i=1,2,....,13,$ are functions of $x$ alone and are connected to the earlier functions through the following relations, namely
\begin{eqnarray}
\hspace{-.2in}&&C_1=B_1(A_1+A_5c_2)+2A_5B_{1x},\,\,\,C_2=B_1(A_2+A_5c_1)+2A_5B_{2x},\nonumber\\
&&C_3=B_1A_3+2A_5(B_2+B_{3x}),\,\,C_4=B_1A_4+2A_5(B_3+B_{4x}),\,\,C_5=2A_5(B_4+B_{5x}),\nonumber\\
&&C_6=2A_5B_5,\,\,\,C_7=B_1A_6+2A_5B_{6x},\,\,\,C_8=B_1A_7+2A_5(B_6+B_{7x}),\nonumber\\
&&C_9=2A_5(B_7+B_{8x}),\,\,\,C_{10}=2A_5B_8,\,\,C_{11}=2B_1+2A_5B_{9x},\nonumber\\
&&C_{12}=2A_5(B_9+B_{10x}),\,\,C_{13}=2A_5B_{10}.\label{ap11}
\end{eqnarray}

We have now two equations with $\eta$, vide Eqs. (\ref{ap8}) and (\ref{ap10}). Eliminating $\eta$ from these two equations we end up with
\begin{eqnarray}
\left(-B_2C_1+B_1C_2\right) \xi_x+\left(-B_3C_1+B_1C_3\right) \xi_{xx}+\left(-B_4C_1+B_1C_4\right) \xi_{xxx}+B_1C_6 \xi_{xxxxx}\nonumber\\
+\left(-B_5C_1+B_1C_5\right) \xi_{xxxx}+\left(-B_6C_1+B_1C_7\right) \xi_{tx}+\left(-B_7C_1+B_1C_8\right) \xi_{txx}\nonumber\\
+\left(-B_8C_1+B_1C_9\right) \xi_{txxx}+B_1C_{10}\xi_{txxxx}+\left(-B_9C_1+B_1C_{11}\right) \xi_{ttx}\nonumber\\
+\left(-B_{10}C_1+B_1C_{12}\right) \xi_{ttxx}+B_1C_{13}\xi_{ttxxx}=0.\label{ap12}
\end{eqnarray}
Substituting the explicit form of $\xi(=b(t)\Im(x) +a(t))$ where $\Im=\int{F(x)dx}$ and $F(x)=e^{\int{fdx}})$ in the above equation (\ref{ap12}), we arrive at
\begin{eqnarray}
\hspace{-.15in}\left(-B_2C_1+B_1C_2\right) bF+\left(-B_3C_1+B_1C_3\right) bfF+\left(-B_5C_1+B_1C_5\right) (f_{xx}+3f_xf+f^3)bF\nonumber\\
+\left(-B_4C_1+B_1C_4\right) (f_x+f^2)bF+B_1C_6 (f_{xxx}+4ff_{xx}+3f_x^2+6f^2f_x+f^4)bF\nonumber\\
+\left(-B_6C_1+B_1C_7\right) \dot{b}F+\left(-B_7C_1+B_1C_8\right) \dot{b}fF+\left(-B_8C_1+B_1C_9\right) (f_x+f^2)\dot{b}F\nonumber\\
+B_1C_{10}(f_{xx}+3f_xf+f^3)\dot{b}F+\left(-B_9C_1+B_1C_{11}\right) \ddot{b}F+\left(-B_{10}C_1+B_1C_{12}\right) \ddot{b}fF\nonumber\\
+B_1C_{13}(f_x+f^2)\ddot{b}F=0,\label{ap13}
\end{eqnarray}
where overdot denotes differentiation with respect to $t$. Note that in (\ref{ap13}) each and every term inside the brackets are functions of $x$ only. With this observation we rewrite Eq. (\ref{ap13}) in a more compact form
\begin{eqnarray}
D_1(x)b(t)+D_2(x)\dot{b}(t)+D_3(x)\ddot{b}(t)=0,\label{ap13.1}
\end{eqnarray}
where $D_1(x),\, D_2(x)$ and $D_3(x)$ are arbitrary functions of $x$ and are defined to be
\begin{eqnarray}
\hspace{-.2in}D_1&=&F \left(B_1 C_2-C_1B_2+f \left(B_1 C_3-C_1B_3\right)+\left(B_1 C_5-C_1B_5\right)\left(f^3+3ff_x+f_{xx}\right)\right.\nonumber\\
&&\left.+ \left(B_1 C_4-C_1B_4\right)\left(f_x+f^2\right)+B_1 C_6\left(f^4+6f^2f_x+3 f_x^2+4ff_{xx}+f_{xxx}\right)\right),\label{ap13.1a}\\
D_2&=&F \left(B_1 C_7-C_1B_6+f\left(B_1C_8-C_1B_7\right)+\left(B_1C_9-C_1B_8\right)\left(f_x+f^2\right)\right.\nonumber\\
&&\left.+B_1C_{10}\left(f^3+3ff_x+f_{xx}\right)\right),\label{ap13.1b}\\
D_3&=&F \left(B_1 C_{11}-C_1B_9+f\left(B_1 C_{12}-C_1B_{10}\right)+B_1 C_{13}\left(f_x+f^2\right)\right).\label{ap13.1c}
\end{eqnarray}

Since  $D_1(x),\, D_2(x)$ and $D_3(x)$ are arbitrary functions involving $x$ alone and do not vanish in general we conclude that the left hand side can become zero if and only if 
\begin{eqnarray}
\ddot{b}=0,\,\,\,\dot{b}=0,\,\,\,b=0.\label{ap14}
\end{eqnarray}
An arbitrary function $b(t)$ which is present in one of the infinitesimals, $\xi$, is vanishing. From this result we conclude that while integrating the determining Eqs. (\ref{a8})-(\ref{a11}) with $L_1\neq0,L_2\neq0$ and $M=0,N=0$ one will always get lesser parameter Lie point symmetries only.

\end{appendix}

\end{document}